\pdfoutput=1
\documentclass[11pt,twoside,a4paper,cmspaper,final,collab]{cms-tdr}
\def\svnVersion{f9088bd}\def\svnDate{2025/02/08}\def\cmsCernNoTag{CERN-EP-2024-161}\def\cmsCernDate{\today}\def\cmsMessage{Published in the Journal of High Energy Physics as \href{http://dx.doi.org/10.1007/JHEP02(2025)036}{\doi{10.1007/JHEP02(2025)036}.}}
\begin{document}\cmsNoteHeader{EXO-21-011}

\newlength\cmsTabSkip\setlength{\cmsTabSkip}{1ex}

\newcommand{\pp}{\ensuremath{\Pp\Pp}\xspace}
\newcommand{\sqrts}[1][13]{\ensuremath{\sqrt{s}=#1\TeV}\xspace}

\newcommand{\hnl}{{\HepParticle{N}{}{}}\xspace}
\newcommand{\mhnl}{\ensuremath{m_\hnl}\xspace}
\newcommand{\tauhnl}{\ensuremath{\PGt_\hnl}\xspace}
\newcommand{\ctauhnl}{\ensuremath{c\tauhnl}\xspace}
\newcommand{\mixpar}{\ensuremath{V_{\Pell\hnl}}\xspace}
\newcommand{\mixpare}{\ensuremath{V_{\Pe\hnl}}\xspace}
\newcommand{\mixparm}{\ensuremath{V_{\PGm\hnl}}\xspace}
\newcommand{\mixparsq}{\ensuremath{\abs{\mixpar}^2}\xspace}
\newcommand{\mixparesq}{\ensuremath{\abs{\mixpare}^2}\xspace}
\newcommand{\mixparmsq}{\ensuremath{\abs{\mixparm}^2}\xspace}
\newcommand{\mixparemsq}{\ensuremath{\abs{\mixpare\mixparm}^2/(\abs{\mixpare}^2+\abs{\mixparm}^2)}\xspace}
\newcommand{\lone}{\ensuremath{\Pell_1}\xspace}
\newcommand{\ltwo}{\ensuremath{\Pell_2}\xspace}

\newcommand{\sigeta}{\ensuremath{\eta}\xspace}
\newcommand{\abseta}{\ensuremath{\abs{\eta}}\xspace}
\newcommand{\dxy}{\ensuremath{d_{xy}}\xspace}
\newcommand{\dz}{\ensuremath{d_z}\xspace}
\newcommand{\dxyz}{\ensuremath{d_{xyz}}\xspace}
\newcommand{\absdxy}{\ensuremath{\abs{\dxy}}\xspace}
\newcommand{\absdz}{\ensuremath{\abs{\dz}}\xspace}
\newcommand{\Irel}{\ensuremath{I_{\text{rel}}}\xspace}

\newcommand{\dR}{\DR\xspace}
\newcommand{\mlonetwo}{\ensuremath{m(\lone\ltwo)}\xspace}
\newcommand{\mmpmmp}{\ensuremath{m(\PGmpm\PGmmp)}\xspace}
\newcommand{\Dphilonetwo}{\ensuremath{\abs{\Delta\phi(\lone,\ltwo)}}\xspace}
\newcommand{\mSV}{\ensuremath{m_{\text{SV}}}\xspace}
\newcommand{\mlSV}{\ensuremath{m(\lone,\text{SV})}\xspace}
\newcommand{\Deltwod}{\ensuremath{\Delta_{\mathrm{2D}}}\xspace}
\newcommand{\Delthreed}{\ensuremath{\Delta_{\mathrm{3D}}}\xspace}
\newcommand{\Lxyz}{\ensuremath{\Delthreed^{\text{gen}}}\xspace}
\newcommand{\PFNthreshold}{\ensuremath{x}\xspace}
\newcommand{\Njets}{\ensuremath{N(\text{jets})}\xspace}

\newcommand{\Cee}{\ensuremath{\Pe\Pe}\xspace}
\newcommand{\Cmm}{\ensuremath{\PGm\PGm}\xspace}
\newcommand{\Cem}{\ensuremath{\Pe\PGm}\xspace}
\newcommand{\Cme}{\ensuremath{\PGm\Pe}\xspace}

\newcommand{\ABCD}[1]{\ensuremath{N_{\text{#1}}}\xspace}

\newcommand{\Wjets}{\ensuremath{\PW\mkern-1mu\text{+jets}}\xspace}
\newcommand{\keVns}{\ensuremath{\text{ke\hspace{-.08em}V}}\xspace}

\cmsNoteHeader{EXO-21-011}
\title{Search for long-lived heavy neutral leptons in proton-proton collision events with a lepton-jet pair associated with a secondary vertex at \texorpdfstring{\sqrts}{sqrt(s)=13 TeV}}
\author{The CMS Collaboration}
\date{\today}

\abstract{A search for long-lived heavy neutral leptons (HNLs) using proton-proton collision data corresponding to an integrated luminosity of 138\fbinv collected at \sqrts with the CMS detector at the CERN LHC is presented. Events are selected with a charged lepton originating from the primary vertex associated with the proton-proton interaction, as well as a second charged lepton and a hadronic jet associated with a secondary vertex that corresponds to the semileptonic decay of a long-lived HNL. No excess of events above the standard model expectation is observed. Exclusion limits at 95\% confidence level are evaluated for HNLs that mix with electron and/or muon neutrinos. Limits are presented in the mass range of 1--16.5\GeV, with excluded square mixing parameter values reaching as low as $2\times10^{-7}$.
For masses above 11\GeV, the presented limits exceed all previous results in the semileptonic decay channel, and for some of the considered scenarios are the strongest to date.}

\hypersetup{%
pdfauthor={CMS Collaboration},%
pdftitle={Search for long-lived heavy neutral leptons in proton-proton collision events with a lepton and jets from a secondary vertex at sqrt(s)=13 TeV},%
pdfsubject={CMS},%
pdfkeywords={CMS, heavy neutrino, displaced}}

\maketitle

\section{Introduction}

The discovery of neutrino oscillations~\cite{Super-Kamiokande:1998kpq, SNO:2002tuh, KamLAND:2002uet} has provided experimental evidence that neutrinos have nonzero masses~\cite{Bilenky:2016pep}.
Constraints from cosmological probes, such as measurements of the cosmic microwave background~\cite{Planck:2018vyg, eBOSS:2020yzd, Sakr:2022ans}, as well as direct measurements~\cite{Formaggio:2021nfz, KATRIN:2021uub}, indicate that the neutrino masses are orders of magnitude smaller than the masses of other fermions in the standard model (SM) of particle physics.
A possible way to generate gauge-invariant mass terms for the SM neutrinos and explain their small mass scale is the see-saw mechanism, which requires the introduction of new heavy states with right-handed chirality~\cite{Minkowski:1977sc, Yanagida:1979as, Gell-Mann:1979vob, Glashow:1979nm, Mohapatra:1979ia, Schechter:1980gr, Shrock:1980ct, Cai:2017mow}.
These new states, referred to as heavy neutral leptons (HNLs), can also provide explanations for other open questions in high-energy physics, such as the nature of dark matter~\cite{Dodelson:1993je, Boyarsky:2018tvu} or the matter-antimatter asymmetry in the universe~\cite{Fukugita:1986hr, Chun:2017spz, Drewes:2021nqr}.

The HNL flavor eigenstates are singlets of the SM gauge groups, and thus do not interact with the SM particles through the electroweak or strong interaction.
They do, however, form mixed mass eigenstates with the SM neutrinos.
This mixing generates HNL-dominated mass eigenstates with effective electroweak couplings, and hence these mass eigenstates can be produced and decay via the electroweak interaction.
For an explanation of both neutrino observations and cosmological problems, an SM extension with HNLs generally needs to provide three HNL generations with couplings to all SM neutrino generations~\cite{Asaka:2005an}.
In the following, we will refer to the HNL mass eigenstates and indicate them with the symbol \hnl.
A wide range of HNL masses \mhnl and mixing parameters \mixpar can yield mass values for the SM neutrinos consistent with observed neutrino oscillations and offer an explanation for the matter-antimatter asymmetry in the universe.
In the case of, \eg, $\mhnl=10\GeV$, these requirements are found to be fulfilled for a range $10^{-11}<\mixparsq<10^{-5}$~\cite{Drewes:2021nqr}; a significant part of which can be probed in proton-proton (\pp) collisions at the CERN LHC.

The experimental signature of HNL models has been studied extensively~\cite{delAguila:2008cj, Atre:2009rg, Tello:2010am, Deppisch:2015qwa, Cai:2017mow, Pascoli:2018heg, Alimena:2019zri, Abdullahi:2022jlv, Antel:2023hkf}, and searches have been conducted at the LHC~\cite{CMS:EXO-11-076, CMS:EXO-12-057, ATLAS:2015gtp, CMS:EXO-14-014, CMS:EXO-17-012, CMS:EXO-17-028, ATLAS:2019kpx, LHCb:2020wxx, CMS:EXO-20-009, ATLAS:2022atq, CMS:EXO-21-003, ATLAS:2023tkz, CMS:EXO-21-013, CMS:EXO-22-017, CMS:EXO-22-011, CMS:EXO-22-019, ATLAS:2024rzi, CMS:EXO-23-006}, at other colliders~\cite{DELPHI:1996qcc, Belle:2013ytx, BaBar:2022cqj, Belle:2022tfo, Belle:2024wyk}, and in fixed-target experiments~\cite{WA66:1985mfx, CHARM:1985nku, NA3:1986ahv, CHARMII:1994jjr, NuTeV:1999kej, T2K:2019jwa, NA62:2020mcv, NA62:2021bji, ArgoNeuT:2021clc, MicroBooNE:2022ctm, MicroBooNE:2023eef}, covering a mass range from a few \keVns to several \TeVns~\cite{Antel:2023hkf}.
The proper lifetime $t$ of HNLs follows an exponential distribution $P(t)=\tauhnl^{-1}\exp(-t/\tauhnl)$ with mean proper lifetime \tauhnl depending on \mhnl and \mixpar as~\cite{Atre:2009rg}
\begin{equation}
    \tauhnl\propto\mhnl^{-5}\abs{\mixpar}^{-2}.
\end{equation}
For $\mhnl<20\GeV$, the value of \tauhnl can become so large that many HNLs will decay after a time $t$ long enough for the HNL decay products to emerge from a secondary vertex (SV) that is spatially displaced from the primary vertex (PV) and separately reconstructable.
Collider searches for HNL production can thus be optimized either for short-lived HNL scenarios with prompt signatures, or for long-lived HNL scenarios with displaced signatures.

\begin{figure}[!t]
\centering
\includegraphics[width=0.45\textwidth]{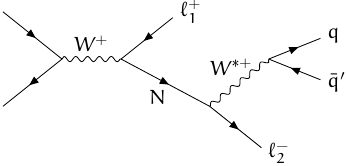}
\hspace{0.05\textwidth}
\includegraphics[width=0.45\textwidth]{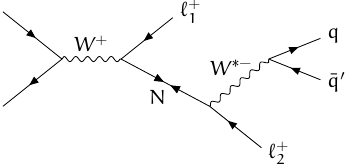}
\caption{%
    Examples of LO Feynman diagrams for production and decay of an HNL (indicated with the symbol \hnl) resulting in a final state with two charged leptons and two quarks.
    In the left diagram, the HNL is a Dirac particle and thus the two charged leptons must have opposite charge.
    In the right diagram, the HNL is a Majorana particle and the two charged leptons can have the same charge.
}
\label{fig:hnldiagram}
\end{figure}

This paper reports a search for \PW boson decays to an HNL and a charged lepton \lone, with a subsequent HNL decay to a second charged lepton \ltwo and two quarks.
The search is optimized for displaced HNL decays with $1<\mhnl<20\GeV$, where the large boost of the HNL in the \PW boson decay causes the two quarks to be merged into a single jet that can also encompass \ltwo.
The experimental signature thus consists of a prompt \lone originating from the PV, and a nonprompt \ltwo encompassed in a jet originating from an SV.
We consider only electrons and muons for \lone and \ltwo, and refer to them as ``leptons'' in the following.
Two possible decay modes are considered for the HNLs: lepton-number-conserving (LNC) decays with \lone and \ltwo of opposite sign (OS), and lepton-number-violating (LNV) decays with same-sign (SS) leptons.
The case where both LNC and LNV decays are possible is referred to as the ``Majorana'' nature of the HNL, whereas the ``Dirac'' nature refers to HNLs with only LNC decays.
As a consequence of having twice the number of possible decays, the decay width (\tauhnl) of a Majorana HNL is twice (half) that of a Dirac HNL with the same \mhnl and \mixparsq.
Examples of leading-order (LO) Feynman diagrams of the studied process with either an LNC or LNV decay are shown in Fig.~\ref{fig:hnldiagram}.

The analyzed \pp collision data were recorded by the CMS experiment at \sqrts in 2016--2018, and correspond to an integrated luminosity of 138\fbinv.
We select events with one prompt lepton and apply a dedicated reconstruction of the SV with an associated nonprompt lepton and a jet.
Events are grouped into the four lepton flavor categories \Cee, \Cmm, \Cem, and \Cme, where the first (second) symbol refers to the flavor of \lone (\ltwo).
A machine-learning algorithm that uses information related to the nonprompt lepton, the jet, and the SV is trained to distinguish between HNL signal and SM background events.
Background contributions from SM processes can arise when a displaced lepton forms an SV by overlapping with a random track, when a photon converts into a lepton pair in the detector material, or when a heavy-flavor hadron decay results in a displaced lepton associated with a genuine SV.
These contributions are estimated from control samples in data.
For the interpretation of our results, we consider simplified models with a single HNL that couples exclusively to electron and/or muon neutrinos (Refs.~\cite{Abada:2018sfh, Tastet:2021vwp, Boiarska:2021yho, Abada:2022wvh} discuss limitations of this approach).
The \Cee (\Cmm) channel is used to constrain \mixparesq (\mixparmsq) in HNL scenarios with an exclusive coupling to electron (muon) neutrinos, whereas the \Cem and \Cme channels are used together to constrain \mixparemsq in HNL scenarios with couplings to both electron and muon neutrinos~\cite{Atre:2009rg}.
Exclusion limits are derived on the respective \mixparsq values as functions of \mhnl.
Tabulated results are provided in the HEPData record for this analysis~\cite{hepdata}.

Recent analyses by the CMS experiment have used the same data set to search for HNLs in various mass ranges and final-state signatures.
The search presented in Ref.~\cite{CMS:EXO-21-013} targets the same decay channel and \mhnl range as the one presented in this paper, but uses a deep-neural-network-based jet tagger to identify the displaced decay products without explicitly reconstructing an SV.
The explicit SV reconstruction employed in the analysis presented here results in more stringent limits in the region with $\mhnl>10\GeV$ and high \tauhnl values.
Searches for HNL production and decay resulting in events with three leptons have been performed in the mass range $10\GeV<\mhnl<1.5\TeV$, where all leptons are required to originate from the PV~\cite{CMS:EXO-22-011}, and in the mass range $1<\mhnl<20\GeV$, where two of the leptons are required to originate from an SV~\cite{CMS:EXO-20-009}.
Other results target the range $1<\mhnl<3\GeV$ by reconstructing HNLs with $\ctauhnl>0.1\unit{m}$ decaying in the muon system~\cite{CMS:EXO-22-017} or by searching for HNLs in \PB meson decays~\cite{CMS:EXO-22-019}.
A summary of all HNL searches performed by the CMS Collaboration is presented in Ref.~\cite{CMS:EXO-23-006}.

\section{The CMS experiment and event reconstruction}
\label{sec:detector}

The central feature of the CMS apparatus is a superconducting solenoid of 6\unit{m} internal diameter, providing a magnetic field of 3.8\unit{T}.
Within the solenoid volume are a silicon pixel and strip tracker, a lead tungstate crystal electromagnetic calorimeter (ECAL), and a brass and scintillator hadron calorimeter (HCAL), each composed of a barrel and two endcap sections.
Forward calorimeters extend the pseudorapidity (\sigeta) coverage provided by the barrel and endcap detectors.
Muons are measured in gas-ionization detectors embedded in the steel flux-return yoke outside the solenoid.
More detailed descriptions of the CMS detector, together with a definition of the coordinate system used and the relevant kinematic variables, can be found in Refs.~\cite{CMS:Detector-2008, CMS:PRF-21-001}.

Events of interest are selected using a two-tiered trigger system.
The first level, composed of custom hardware processors, uses information from the calorimeters and muon detectors to select events at a rate of around 100\unit{kHz} within a fixed latency of about 4\mus~\cite{CMS:TRG-17-001}.
The second level, known as the high-level trigger (HLT), consists of a farm of processors running a version of the full event reconstruction software optimized for fast processing, and reduces the event rate to around 1\unit{kHz} before data storage~\cite{CMS:TRG-12-001, CMS:TRG-19-001}.

Charged-particle tracks are reconstructed across the entire \sigeta range of the silicon tracker, which is $\abseta<2.5$ with the tracker used in 2016 and $\abseta<3.0$ with the new pixel detector installed at the start of 2017~\cite{CMS:Pixel-2021}.
Track reconstruction enables the detection of charged particles with transverse momentum (\pt) as low as 0.1\GeV and originating from distances up to 60\unit{cm} from the beam line.
In simulation with 2017 detector conditions, the tracking efficiency for charged particles with $\pt>0.9\GeV$ of better than 90, 60, and 25\% for tracks originating at distances smaller than 2, 10, and 48\unit{cm} from the beamline, respectively, is found~\cite{CMS:DP-2017-015}.
These charged-particle tracks are then clustered to form vertices.
To reconstruct vertices corresponding to \pp interactions, the clustering is performed based on the $z$ coordinate of the closest approach of the tracks to the beam line.
The vertex position is estimated with an adaptive vertex fitter~\cite{Fruhwirth:2007hz} using the collection of tracks compatible with originating from the same interaction.
From the vertices reconstructed in this way, the PV is taken to be the one corresponding to the hardest scattering in the event, evaluated using tracking information alone, as described in Section 9.4.1 of Ref.~\cite{CMS:TDR-15-02}.

The impact parameter of a track is defined as the distance of the track from the PV at its point of closest approach.
The three-dimensional (3D) impact parameter is referred to as \dxyz, the two-dimensional (2D) value in the plane transverse to the beam line as \dxy, and the longitudinal value along the beam line as \dz.
For nonisolated particles of $1<\pt<10\GeV$ and $\abseta<1.4$, the track resolutions in 2016 were typically 1.5\% in \pt and 25--90 (45--150)\mum in \dxy (\dz)~\cite{CMS:TRK-11-001}.
From 2017 onward, the upgraded tracker measured particles up to $\abseta<3.0$ with typical resolutions of 1.5\% in \pt and 20--75\mum in \dxy for nonisolated particles of $1<\pt<10\GeV$~\cite{CMS:DP-2020-049}.

To reconstruct vertices corresponding to the decay of long-lived particles displaced from the PV, we employ a modified version of the inclusive vertex finder (IVF) algorithm~\cite{CMS:BPH-10-010, CMS:BTV-16-002} originally developed for the identification of displaced \PB hadron decays.
The IVF clusters tracks that have at least six hits in the tracker and $\pt>0.8\GeV$, and is seeded with tracks that additionally have a 3D impact parameter significance, defined as \dxyz divided by its uncertainty, above 1.2.
The 3D (2D) distance of the SV to the PV is referred to as \Delthreed (\Deltwod).
All SVs are retained that have a significance of \Delthreed (\Deltwod) above 0.5 (2.5), while no explicit requirements are applied on the values of \Delthreed or \Deltwod directly.
From the high number of SVs reconstructed with the IVF, we select the candidate vertex corresponding to the HNL decay by requiring that the SV contains the track associated with \ltwo, selected as described in Section~\ref{sec:eventselection}.
If more than one SV contains the \ltwo track, we select the SV with the highest associated \pt.

The particle-flow (PF) algorithm~\cite{CMS:PRF-14-001} aims to reconstruct and identify each individual particle in an event, with an optimized combination of information from the various elements of the CMS detector.
The energy of photons is obtained from the ECAL measurement.
The energy of electrons is determined from a combination of the electron momentum at the PV, the energy of the corresponding ECAL cluster, and the energy sum of all bremsstrahlung photons spatially compatible with originating from the electron track.
The energy of muons is obtained from the curvature of the corresponding track.
The energy of charged hadrons is determined from a combination of their momentum measured in the tracker and the matching ECAL and HCAL energy deposits, corrected for the response function of the calorimeters to hadronic showers.
Finally, the energy of neutral hadrons is obtained from the corresponding corrected ECAL and HCAL energies.

Reconstructed PF particles are clustered into jets with the anti-\kt algorithm~\cite{Cacciari:2008gp, Cacciari:2011ma} using a distance parameter of 0.4.
The jet momentum is determined as the vectorial sum of all particle momenta in the jet, and is found from simulation to be, on average, within 5--10\% of the true momentum over the whole \pt spectrum and detector acceptance.
Additional \pp interactions within the same or nearby bunch crossings (pileup) can contribute with additional tracks and calorimetric energy depositions, increasing the apparent jet momentum.
To mitigate this effect, tracks identified to be originating from pileup vertices are discarded and an offset correction is applied to account for remaining contributions~\cite{CMS:JME-18-001}.
Jet energy corrections are derived from simulation studies so that the average measured energy of jets becomes identical to that of particle-level jets.
In situ measurements of the momentum balance in dijet, photon+jet, \PZ{+}jet, and multijet events are used to determine any residual differences between the jet energy scale in data and simulation, and appropriate corrections are made~\cite{CMS:JME-13-004}.
Additional selection criteria are applied to each jet to remove jets potentially dominated by instrumental effects or reconstruction failures~\cite{CMS:JME-18-001}.
Finally, jets are required to have $\pt>20\GeV$ and $\abseta<2.4$.

Electrons are reconstructed within the geometrical acceptance of the silicon strip tracker, which is $\abseta<2.5$.
The electron momentum is estimated by combining the energy measurement in the ECAL with the momentum measurement in the tracker.
The momentum resolution for electrons with $\pt\approx45\GeV$ from $\PZ\to\Cee$ decays ranges from 1.6 to 5\%.
It is generally better in the barrel region than in the endcaps, and also depends on the bremsstrahlung energy emitted by the electron as it traverses the material in front of the ECAL~\cite{CMS:EGM-17-001, CMS:DP-2020-021}.

Muons are measured in the range $\abseta<2.4$, with detection planes made using three technologies: drift tubes, cathode strip chambers, and resistive plate chambers.
The efficiency to reconstruct and identify muons is greater than 96\%. Matching muons to tracks measured in the silicon tracker results in a relative \pt resolution, for muons with \pt up to 100\GeV, of 1\% in the barrel and 3\% in the endcaps~\cite{CMS:MUO-16-001}.

\section{Data and simulated samples}
\label{sec:samples}

{\tolerance=800
Data events have been recorded with single-electron and single-muon triggers.
For the single-electron trigger, the presence of at least one isolated electron at the HLT is required with a \pt threshold of 27 (32)\GeV in 2016 (2017--2018)~\cite{CMS:DP-2020-016}.
For the single-muon trigger, the presence of at least one isolated muon at the HLT is required with a \pt threshold of 24\GeV throughout all data-taking years~\cite{CMS:MUO-19-001}.
\par}

Monte Carlo (MC) event simulation is used to generate HNL signal events for the evaluation of the signal selection efficiency, and SM background events for the validation of the background estimation.
For each signal and background process, separate simulated event samples are generated corresponding to the conditions of the three data-taking years.
For the 2016 (2017--2018) samples, the NNPDF 3.0~\cite{NNPDF:2014otw} (3.1~\cite{NNPDF:2017mvq}) parton distribution functions (PDFs) are used in the simulation of the hard process.
All generators are interfaced with the \PYTHIA~8.226 (8.230) simulation~\cite{Sjostrand:2014zea} for parton showering and hadronization, using the CUETP8M1~\cite{CMS:GEN-14-001} (CP5~\cite{CMS:GEN-17-001}) tune for the underlying event modeling in case of the event modeling for the 2016 (2017--2018) samples.
Simulated minimum bias events are superimposed on the generated events to reproduce the pileup distribution obtained from the measured instantaneous luminosity.
For all simulated events, the CMS detector response is modeled with the \GEANTfour toolkit~\cite{GEANT4:2002zbu}.

Background samples for \PW boson production in association with jets (\Wjets) are generated at next-to-LO (NLO) with the \MGvATNLO~2.4.2 event generator~\cite{Alwall:2014hca}, including up to two additional partons in the matrix-element calculation.
For Drell--Yan (DY) dilepton production, event samples are generated at LO (NLO) with up to four (two) additional partons in the matrix-element calculation with \MGvATNLO.
Top quark (pair and single production) and diboson ($\PW\PW$, $\PW\PZ$, $\PZ\PZ$, $\PW\PGg$, $\PZ\PGg$) processes are simulated with the \POWHEG~v2~\cite{Nason:2004rx, Frixione:2007nw, Frixione:2007vw, Alioli:2009je, Alioli:2010xd, Re:2010bp, Melia:2011tj, Nason:2013ydw} and \MGvATNLO event generators at NLO.
Samples generated with \MGvATNLO at LO (NLO) use the MLM~\cite{Alwall:2007fs} (FxFx~\cite{Frederix:2012ps}) merging algorithm.
For the background from events comprised uniquely of jets produced through the strong interaction (QCD multijet events), event samples are generated at LO with \PYTHIA.

Signal events for HNL production are simulated at LO with \MGvATNLO, using a model that extends the SM particle content by introducing up to three right-handed neutrinos~\cite{Atre:2009rg, Alva:2014gxa, Degrande:2016aje}.
Event samples are generated for a single HNL mass eigenstate with pure electron neutrino couplings, pure muon neutrino couplings, and equal electron and muon neutrino couplings, covering the parameter ranges $1<\mhnl<20\GeV$ and $10^{-9}<\mixparsq<10^{-1}$.
The cross section of the HNL event samples is calculated at LO with \MGvATNLO, and scaled to next-to-NLO (NNLO) precision using a $K$-factor of 1.089 derived from the comparison of SM \PW boson production samples generated at LO accuracy with \MGvATNLO, using the same settings as the signal simulation, and the cross section calculated at NNLO accuracy with the \textsc{fewz}~v3.1 program~\cite{Melnikov:2006kv, Gavin:2010az, Gavin:2012sy, Li:2012wna}.

For a fixed value of \mhnl, we emulate an event sample with a different \mixparsq value by applying a global weight to match the expected cross section and a per-event weight to correct the proper lifetime distribution.
The cross section is proportional to \mixparsq~\cite{Degrande:2016aje} and the global weight is accordingly calculated as the ratio of the new to the simulated \mixparsq value.
To reweight from \tauhnl values $\tau_0$ to $\tau_1$, the per-event weights for events with proper lifetime $t$ are calculated as
\begin{equation}\label{eq:lifetimeweights}
    W(t)=\frac{P(t;\tau_1)}{P(t;\tau_0)}
    =\frac{\tau_0}{\tau_1}\exp\bigg[-t\bigg(\frac{1}{\tau_1}-\frac{1}{\tau_0}\bigg)\bigg].
\end{equation}
The simulation is performed assuming Majorana HNLs, \ie, both LNC and LNV decays are included.
To obtain an event sample for Dirac HNLs, we select only events with LNC decays.
The production cross section is the same for Majorana and Dirac HNLs, and thus the global weight is only adjusted for the smaller number of generated events in the sample.
The per-event weights are calculated with Eq.~\eqref{eq:lifetimeweights} but using that $\tau_1$ is twice as large for a Dirac HNL with respect to a Majorana HNL of same \mhnl and \tauhnl.

\section{Event selection}
\label{sec:eventselection}

To select prompt and nonprompt leptons and to reduce the contribution from particles misidentified as leptons, additional identification (ID) criteria are applied to the reconstructed electrons and muons.
We employ two standard sets of electron ID criteria defined in Ref.~\cite{CMS:EGM-17-001} that both have an efficiency of 90\%, where the first is based on sequential selection of requirements on seven observables (``sequential ID''), and the second is based on a multivariate ID discriminant trained with an extended set of observables (``multivariate ID'').
For muons, we use the set of ``medium'' ID criteria defined in Ref.~\cite{CMS:MUO-16-001} that has an efficiency of 99.5\%.
In addition, the relative isolation \Irel, defined as the scalar sum of the of surrounding particles divided by the lepton \pt, is used to quantify how isolated a reconstructed lepton is from other particles in the event.
The isolation sum requires particles to be within a cone of $\dR=\sqrt{\smash[b]{(\Delta\eta)^2+(\Delta\phi)^2}}<0.3$, where $\Delta\eta$ and $\Delta\phi$ are the \sigeta and azimuthal angle differences between the lepton and the particle, respectively.
Charged hadrons are included only if they originate from the PV, while all neutral hadrons and photons within the cone are included and additional corrections for the contribution of neutral particles from pileup interactions are applied~\cite{CMS:EGM-17-001, CMS:MUO-16-001}.
Finally, we employ selection criteria for the impact parameters of the tracks associated with the electrons and muons, as defined in Section~\ref{sec:detector}.

\begin{table}[!ht]
\centering
\topcaption{%
    Selection criteria for electrons and muons.
    Numbers in parentheses indicate values applied in the 2017--2018 data sets, when different from those for 2016.
}
\label{tab:objectselection}
\renewcommand\arraystretch{1.2}
\begin{tabular}{lcccc}
    \multirow{2}{*}{Criterion} & \multicolumn{2}{c}{Prompt} & \multicolumn{2}{c}{Nonprompt} \\[-2pt]
    & Electrons & Muons & Electrons & Muons \\ \hline
    \pt [{\GeVns}] & $>$30 (34) & $>$30 & $>$7 & $>$7 \\
    \abseta & $<$2.5 & $<$2.4 & $<$2.5 & $<$2.4 \\
    \Irel & $<$0.1 & $<$0.1 & \NA & \NA \\
    \absdxy [cm] & $<$0.02 & $<$0.01 & $>$0.02 & $>$0.02 \\
    \absdz [cm] & $<$0.04 & $<$0.1 & $<$10 & $<$10 \\
    ID & full & full & reduced & reduced \\
\end{tabular}
\end{table}

The selection criteria for prompt and nonprompt electrons and muons are listed in Table~\ref{tab:objectselection}.
Prompt electrons and muons are required to have a \pt above a threshold chosen larger than the HLT single-lepton trigger thresholds, to be isolated, and to originate from the PV.
Nonprompt electrons and muons are selected with a much lower \pt threshold and no isolation requirement.
Instead of consistency with the PV, a minimal requirement on \dxy is imposed, and a loose maximal requirement on \dz is used to reduce background contributions from cosmic ray muons and pileup.
In addition, we require that electrons and muons pass a set of ID criteria as introduced before, where prompt leptons have to pass the ``full'' set and nonprompt leptons only a ``reduced'' set without criteria that are inefficient for displaced leptons.
For prompt electrons, the multivariate ID is used.
For nonprompt electrons, the sequential ID is used except for the requirements on the maximum number of missing hits in the pixel detector and the veto on photon conversions, as defined in Ref.~\cite{CMS:EGM-17-001}.
We use the medium ID for both prompt and nonprompt muons, but remove the requirements on the fraction of valid tracker hits associated with the muon track and on the fit quality of the global muon track fit~\cite{CMS:MUO-16-001} for nonprompt muons.

\begin{figure}[!t]
\centering
\includegraphics[width=0.45\textwidth]{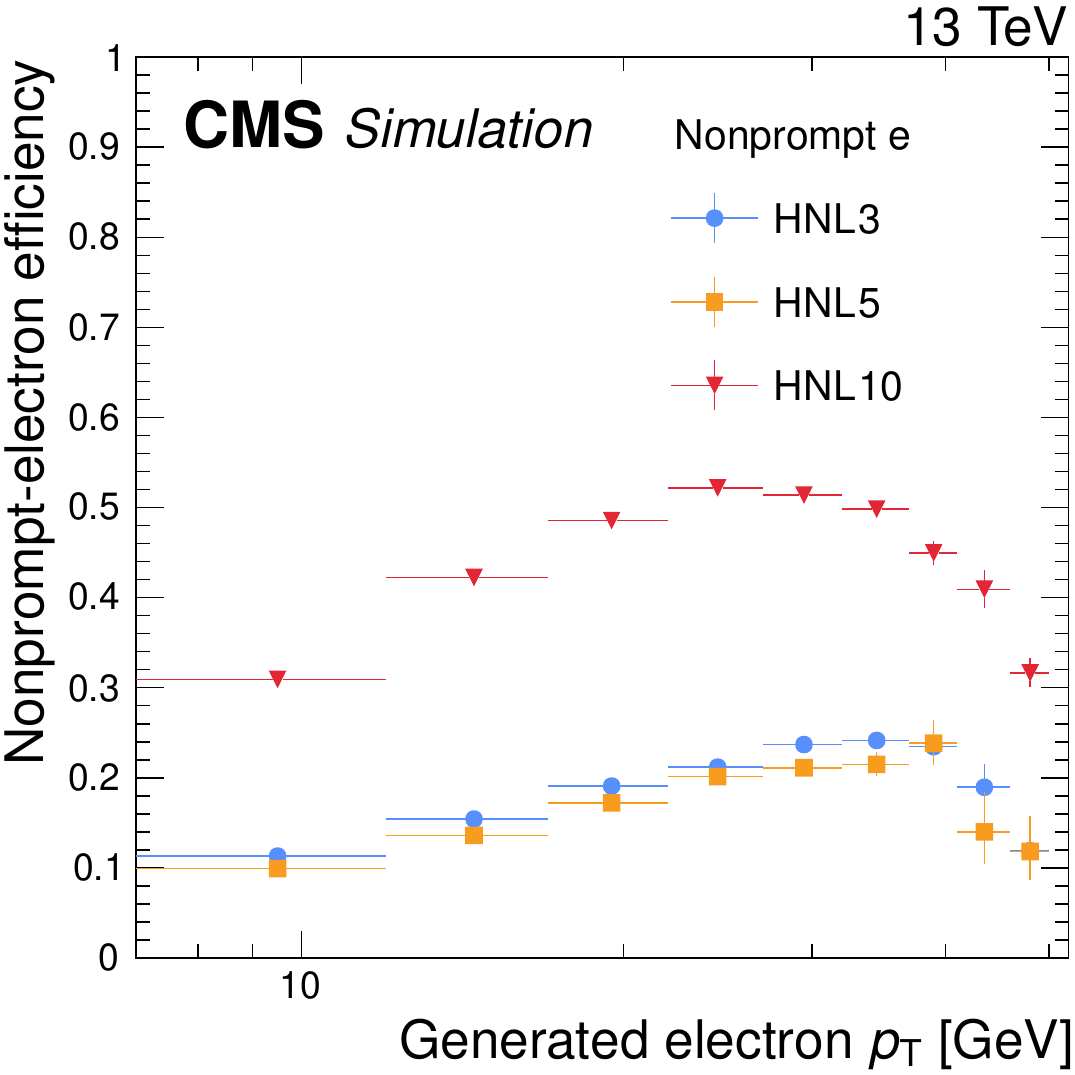}%
\hspace{0.05\textwidth}%
\includegraphics[width=0.45\textwidth]{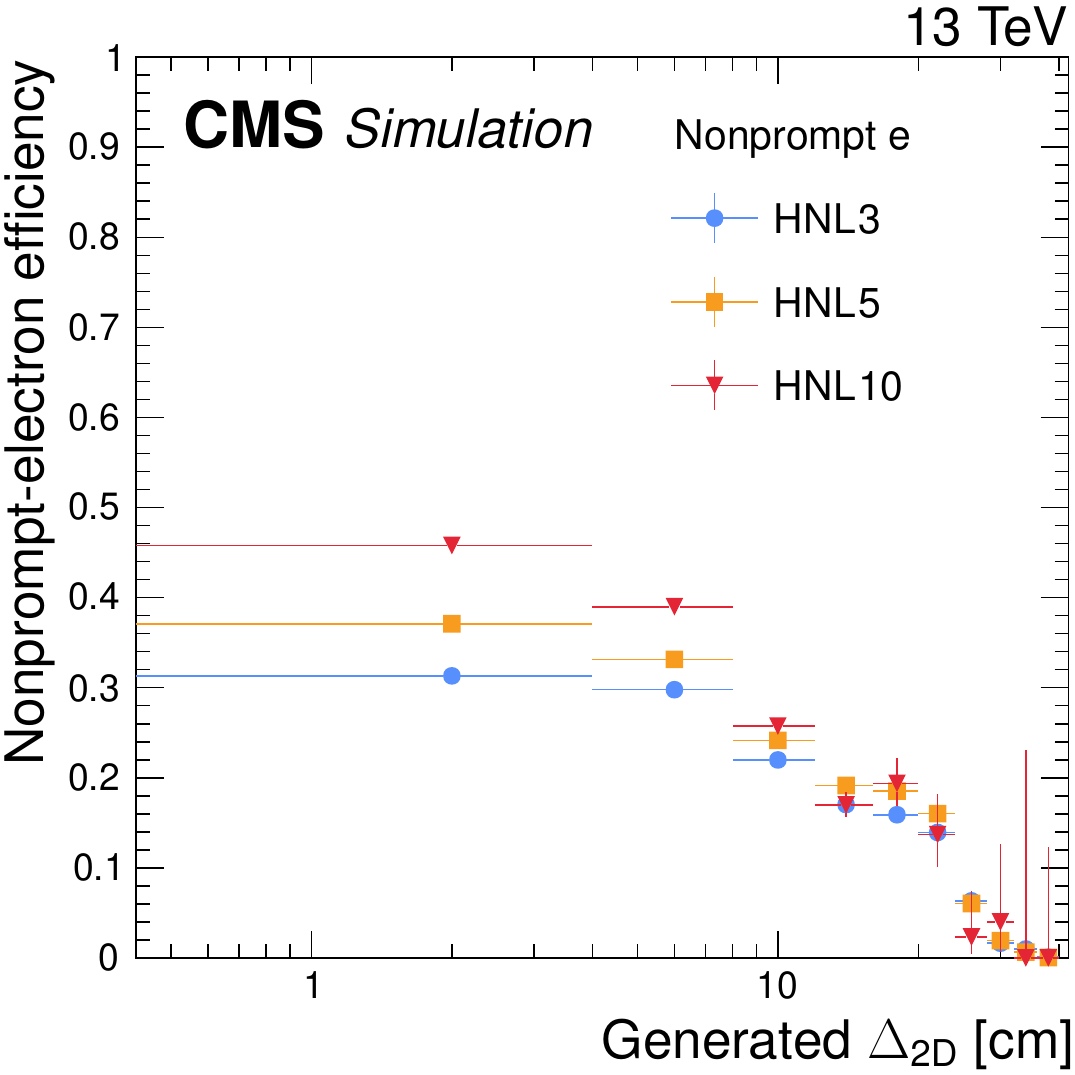}%
\\[1ex]
\includegraphics[width=0.45\textwidth]{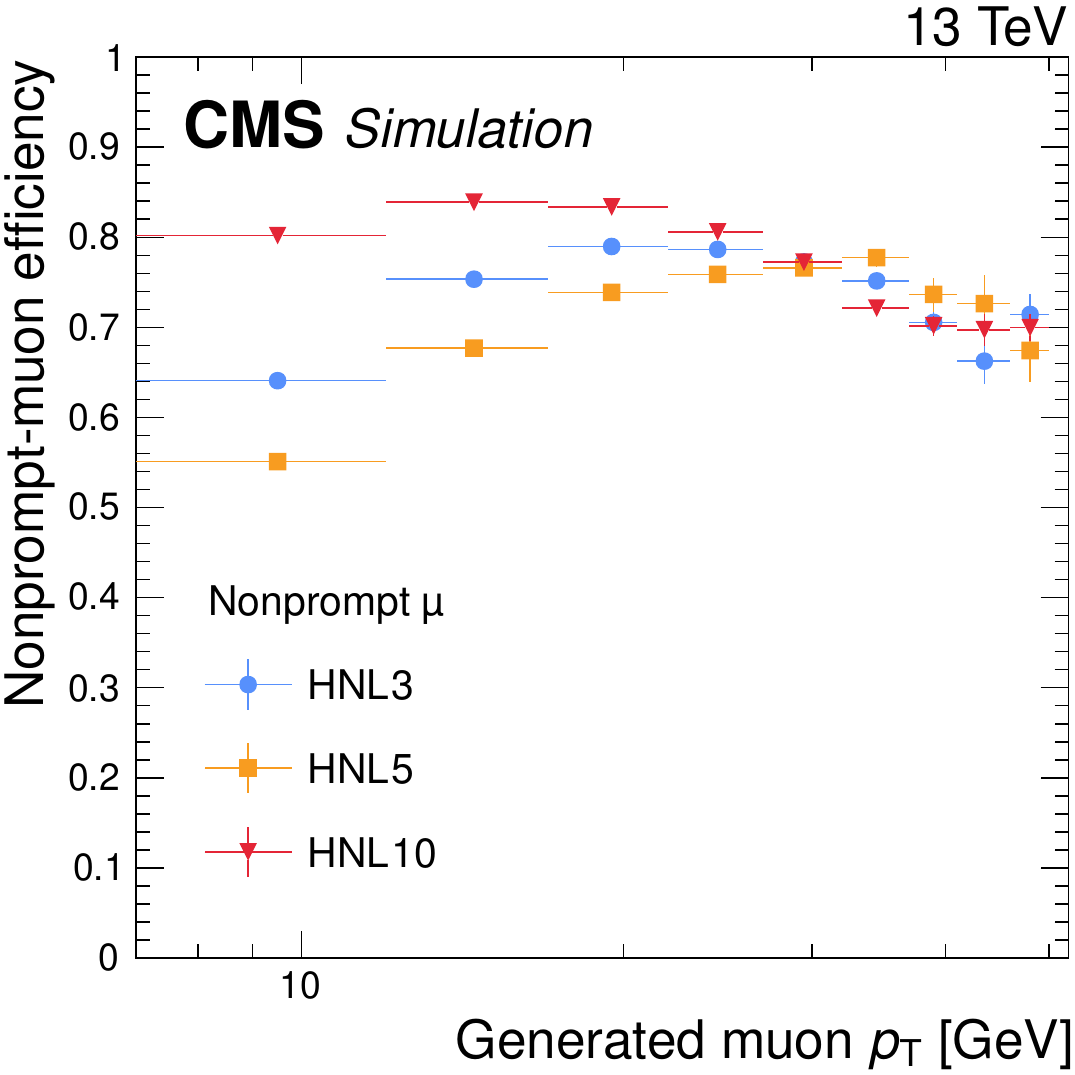}%
\hspace{0.05\textwidth}%
\includegraphics[width=0.45\textwidth]{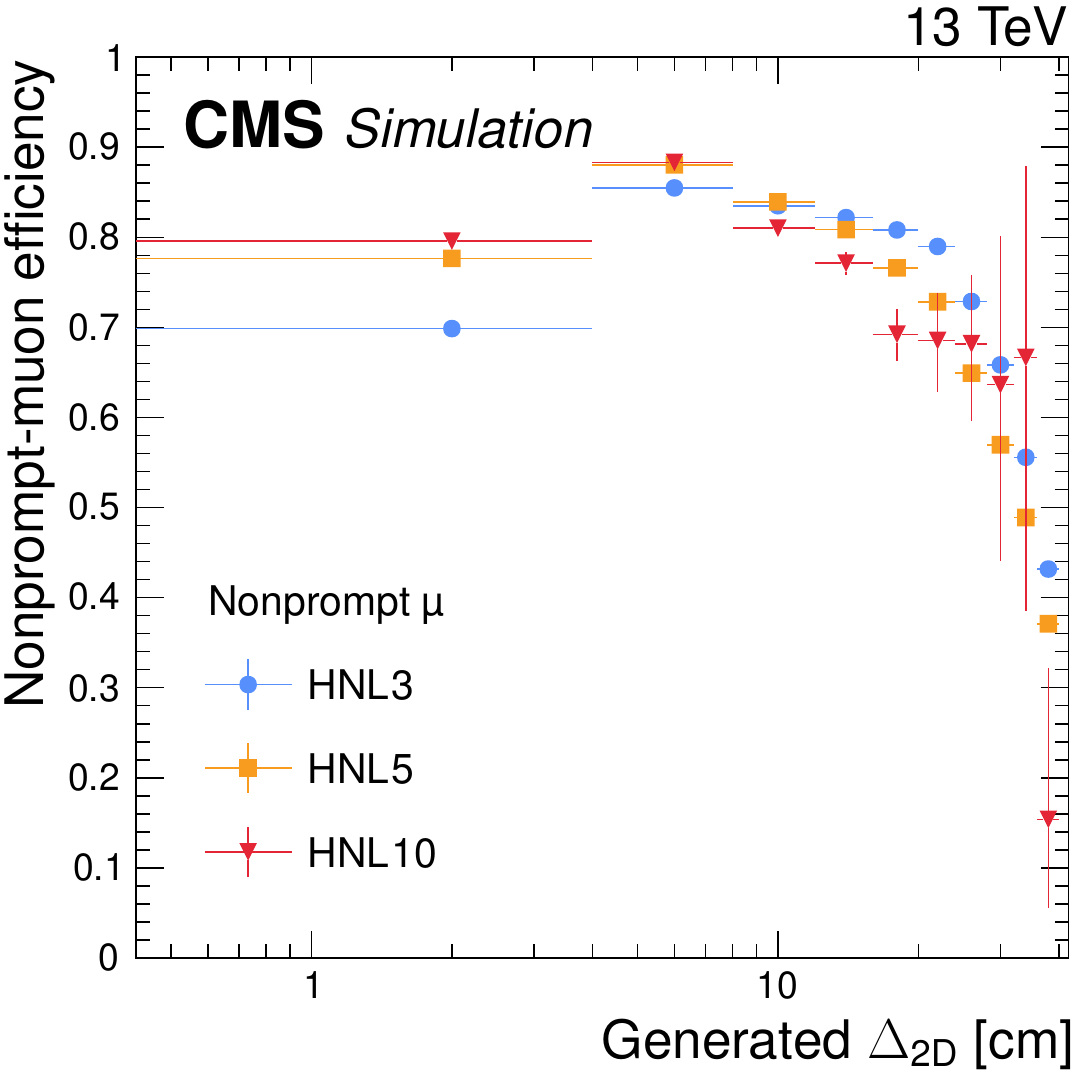}%
\caption{%
    Selection efficiencies of nonprompt electrons (upper) and muons (lower), evaluated in simulated HNL signal events as functions of the generated lepton \pt (left) and transverse displacement of the generated SV (right).
    Three HNL signal scenarios are shown, with $\mhnl=3\GeV$ and $\mixparsq=9.9\times10^{-5}$ (HNL3, corresponding to $\ctauhnl=23\unit{mm}$), $\mhnl=5\GeV$ and $\mixparsq=1.6\times10^{-6}$ (HNL5, corresponding to $\ctauhnl=92\unit{mm}$), and $\mhnl=10\GeV$ and $\mixparsq=5.7\times10^{-7}$ (HNL10, corresponding to $\ctauhnl=7\unit{mm}$).
    The error bars in the plot represent statistical uncertainties.
}
\label{fig:nonpromptefficiency}
\end{figure}

The selection efficiencies for nonprompt electrons and muons are evaluated in HNL signal samples with different $(\mhnl, \mixparsq)$ scenarios, representing \ctauhnl values in the range 7--92\unit{mm}, and shown in Fig.~\ref{fig:nonpromptefficiency} as functions of the generated \pt and the transverse displacement \Deltwod of the generated SV.
The simulated HNL signal events used for this evaluation are selected by requiring a generated nonprompt electron (muon) with $\pt>7\GeV$ and $\abseta<2.5$ (2.4), and a generated HNL decay vertex with $\Deltwod<50\cm$.
The obtained efficiencies reflect various effects in the electron and muon reconstruction methods and identification criteria, as well as effects caused by different \mhnl and \ctauhnl values.
Overall, the better resolution and higher efficiency in the basic reconstruction of muons compared to electrons~\cite{CMS:EGM-17-001, CMS:MUO-16-001} results in higher selection efficiencies for nonprompt muons when compared to electrons, which leads to a better sensitivity of the analysis to \mixparmsq than \mixparesq.
At low values of lepton \pt, the track reconstruction in the inner tracker has a worse efficiency and resolution, decreasing especially the efficiency for nonprompt electrons.
At high values of lepton \pt, the decay products of the HNL will be boosted and thus the nonprompt lepton is closer to or fully included in the jet from the HNL decay, which reduces the efficiency to pass selection criteria designed to reject hadrons misidentified as leptons.
Due to differences in the track reconstruction and combination with ECAL deposits or the muon system track, respectively, the \pt-dependent shape is different between nonprompt electrons and muons.
The selection efficiency of nonprompt electrons decreases quickly with larger displacement, since the dedicated electron track reconstruction is very sensitive to missing hits in the inner tracker~\cite{CMS:EGM-17-001}, whereas the muon reconstruction including the muon system track is less dependent on a well-reconstructed track in the inner tracker.
For nonprompt muons, a small inefficiency can be seen in the first bin of the efficiency as a function of \Deltwod, caused by the minimal requirement on \absdxy listed in Table~\ref{tab:objectselection}.

\begin{figure}[!ht]
\centering
\includegraphics[width=0.45\textwidth]{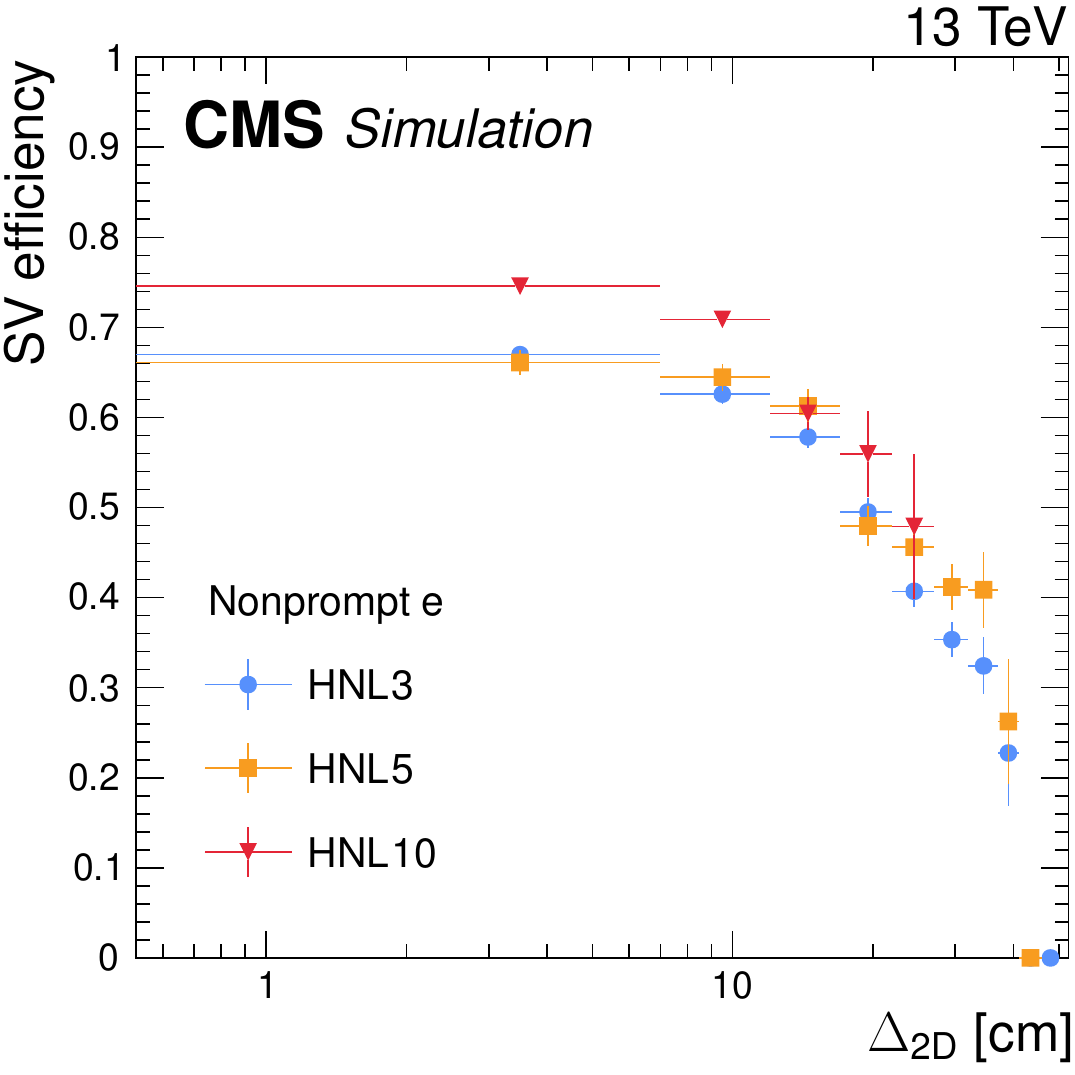}%
\hspace{0.05\textwidth}%
\includegraphics[width=0.45\textwidth]{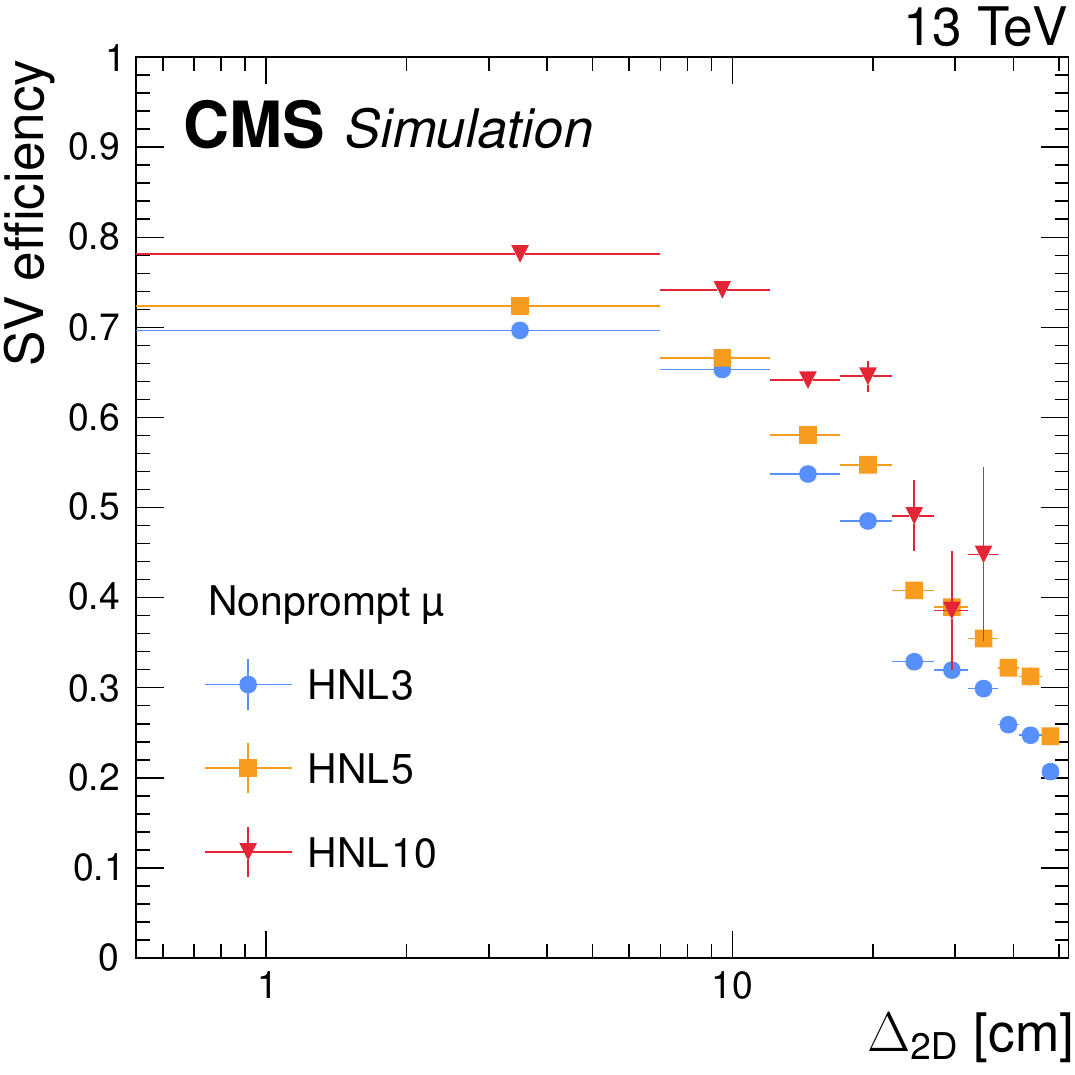}%
\caption{%
    The SV reconstruction efficiency in simulated HNL signal events as a function of the SV displacement for vertices with a nonprompt electron (left) or muon (right).
    Three HNL signal scenarios are shown, with $\mhnl=3\GeV$ and $\mixparsq=9.9\times10^{-5}$ (HNL3, corresponding to $\ctauhnl=23\unit{mm}$), $\mhnl=5\GeV$ and $\mixparsq=1.6\times10^{-6}$ (HNL5, corresponding to $\ctauhnl=92\unit{mm}$), and $\mhnl=10\GeV$ and $\mixparsq=5.7\times10^{-7}$ (HNL10, corresponding to $\ctauhnl=7\unit{mm}$).
    The error bars in the plot represent statistical uncertainties.
}
\label{fig:hnldecaylength}
\end{figure}

Events are selected with exactly one prompt electron or muon (\lone), which is required to match geometrically to the lepton reconstructed at the HLT that satisfies the trigger selection.
Further, events are required to have exactly one nonprompt electron or muon (\ltwo), as well as one jet matched to \ltwo by the requirement $\dR<0.7$.
To reconstruct the HNL decay vertex, the \ltwo track is required to be part of an SV reconstructed by the IVF, as described in Section~\ref{sec:detector}.
The SV reconstruction is validated in simulation by comparing properties of the reconstructed SV with the generated HNL decay vertex.
In Fig.~\ref{fig:hnldecaylength}, the SV reconstruction efficiency is shown as a function of \Deltwod, using HNL signal events with an SV that contains the nonprompt-lepton track and is matched to the generated HNL decay vertex.
The vertex matching is defined by the criterion $\abs{\Delthreed-\Lxyz}/\Lxyz<0.1$, where \Lxyz is the 3D displacement of the generated HNL decay vertex.

Further requirements are imposed on the invariant mass of the two selected leptons, $\mlonetwo>10\GeV$, and on their angular separation, $\Dphilonetwo>0.4$.
These requirements remove poorly reconstructed events, as well as background events originating from low-mass resonances.
Events with at least one additional jet separated from \lone by $\DR>0.4$, as expected for top quark production, are removed and used as an orthogonal event selection for the validation of the background prediction described in Section~\ref{sec:background}.
Events where \lone and \ltwo are OS muons and have $85<\mmpmmp<95\GeV$ are removed to suppress DY dimuon background contributions.
This DY veto is introduced because of a correlation between the observables used in the background estimation observed in the OS dimuon channels, as described in Section~\ref{sec:background}, but is not required for the other channels where no such correlation is observed.
The event selection criteria are summarized in Table~\ref{tab:eventselection}.

\begin{table}[!ht]
\centering
\topcaption{%
    Summary of the event selection criteria.
}
\label{tab:eventselection}
\renewcommand\arraystretch{1.2}
\begin{tabular}{r@{\,}c@{\,}l}
    \multicolumn{3}{c}{Event selection criteria} \\ \hline
    $N(\text{prompt \lone})$ & $=$ & 1 \\
    $N(\text{nonprompt \ltwo})$ & = & 1 \\
    \Njets & = & 1 \\
    $\dR(\ltwo,\text{jet})$ & $<$ & 0.7 \\
    \ltwo & $\in$ & SV \\
    \mlonetwo & $>$ & 10\GeV \\
    \Dphilonetwo & $>$ & 0.4 \\
    \mmpmmp & $\notin$ & [85,95]\GeV \\
\end{tabular}
\end{table}

\begin{figure}[!ht]
\centering
\includegraphics[width=0.42\textwidth]{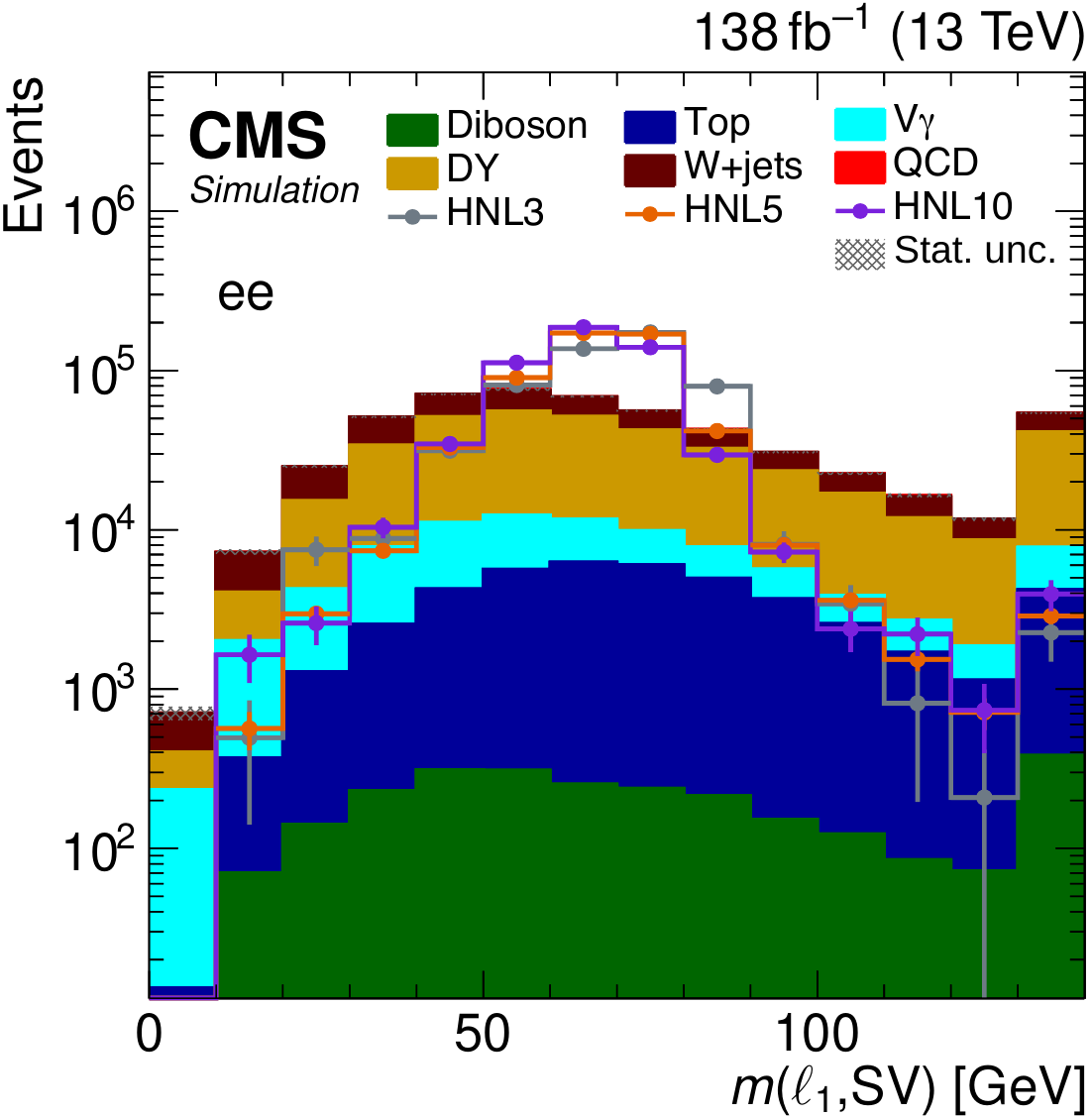}%
\hspace{0.05\textwidth}%
\includegraphics[width=0.42\textwidth]{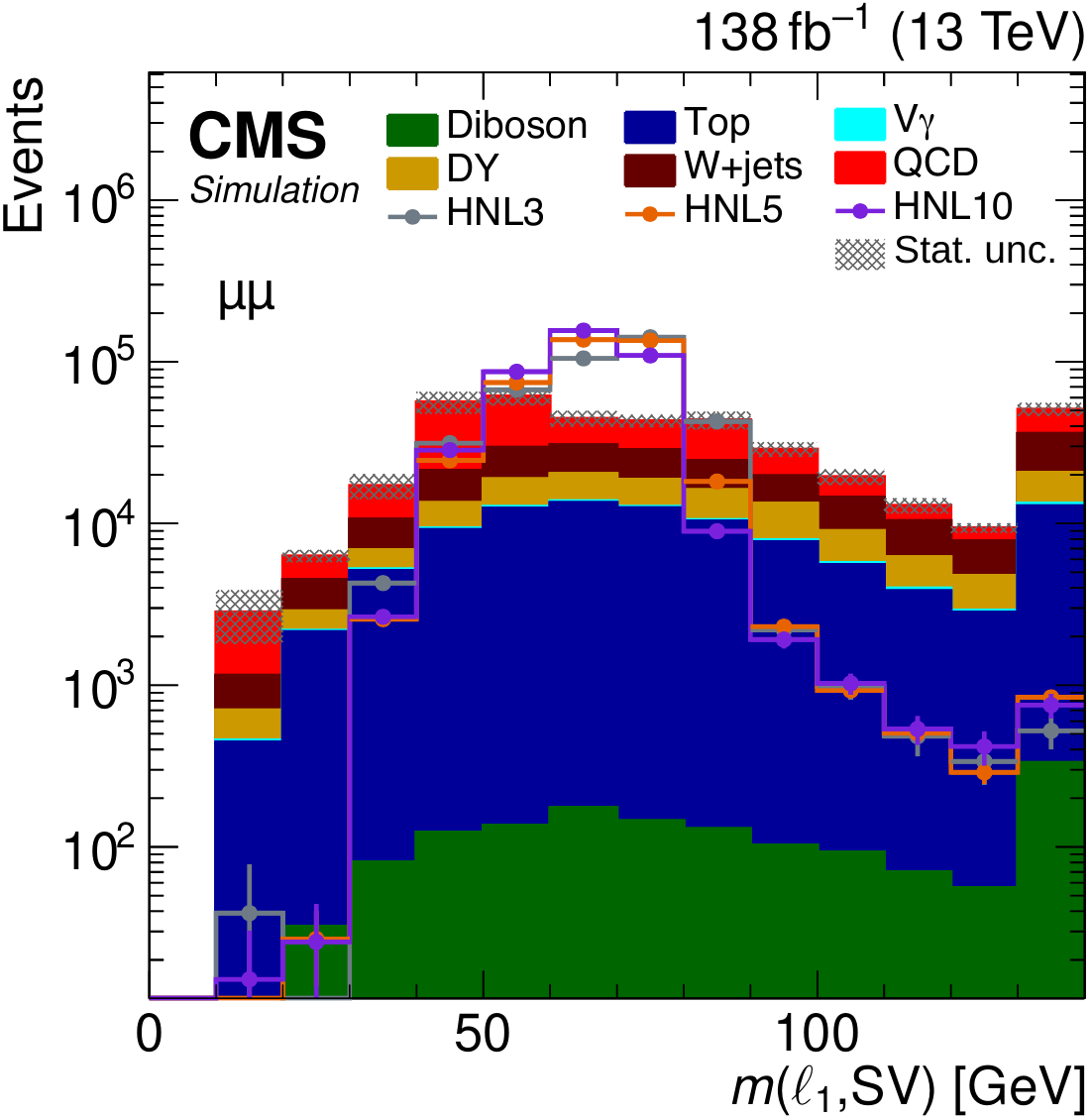} \\[1ex]
\includegraphics[width=0.42\textwidth]{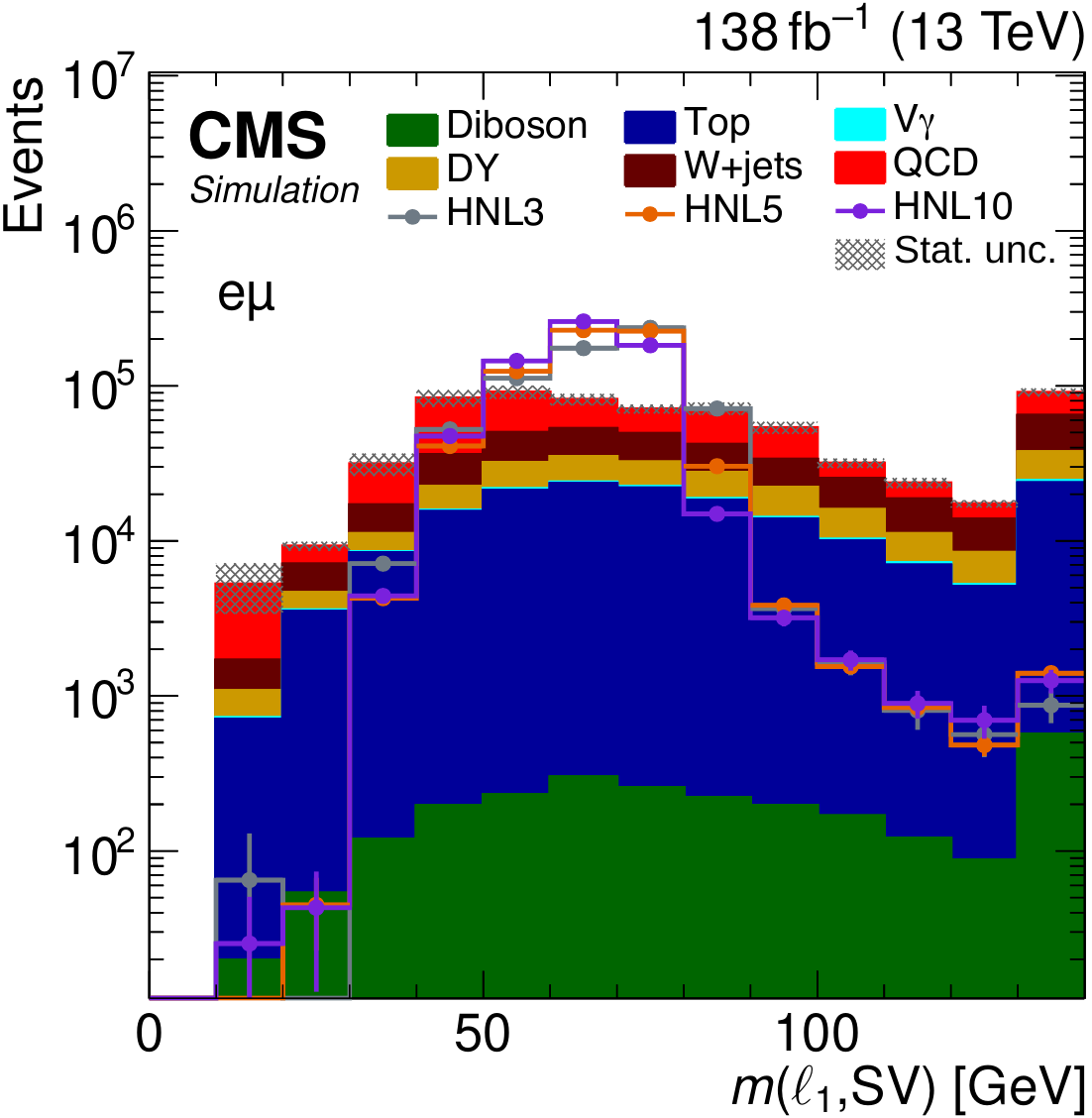}%
\hspace{0.05\textwidth}%
\includegraphics[width=0.42\textwidth]{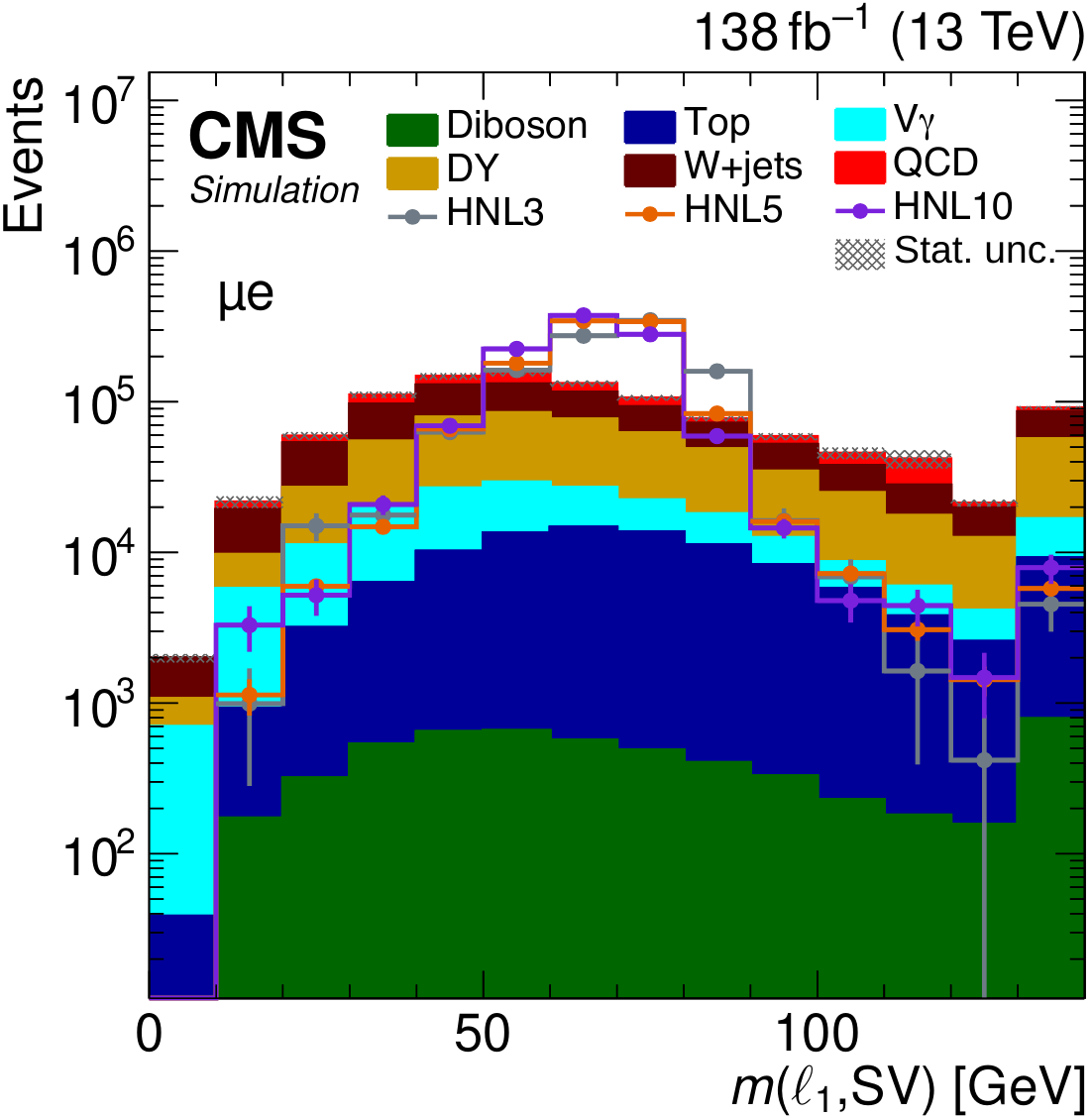}
\caption{%
    The \mlSV distribution of predicted events yields after applying the selection summarized in Table~\ref{tab:objectselection}, for the \Cee (upper left), \Cmm (upper right), \Cem (lower left), and \Cme (lower right) categories.
    The filled histograms display the predicted background yields.
    The lines show the predicted yields for three HNL signal scenarios, with $\mhnl=3\GeV$ and $\mixparsq=9.9\times10^{-5}$ (HNL3), $\mhnl=5\GeV$ and $\mixparsq=1.6\times10^{-5}$ (HNL5), and $\mhnl=10\GeV$ and $\mixparsq=5.7\times10^{-7}$ (HNL10).
    The HNL signal yield is normalized to the total background yield.
    The last bins include the overflow.
}
\label{fig:mlsv}
\end{figure}

Separate event categories are defined based on the flavors of \lone and \ltwo (\Cee, \Cem, \Cme, and \Cmm), and for the cases of SS and OS charges.
The invariant mass of \lone and the tracks belonging to the selected SV, \mlSV, is an important observable for the discrimination between signal and background events.
For HNL signal events, \mlSV is kinematically limited by the \PW boson mass since it includes all charged particles originating from the \PW boson decay in the hard interaction but does not account for potential neutral hadrons and photons formed in the hadronic decay.
The distribution of \mlSV predicted from simulated event samples is shown in Fig.~\ref{fig:mlsv}, separately for the different event categories.

\section{Identification of the SV from an HNL decay}
\label{sec:PFN}

A multivariate discriminant based on machine learning is employed to distinguish between events with a reconstructed SV originating from an HNL decay and background events where, \eg, the jet accidentally overlaps with the nonprompt lepton.
We employ a particle-flow network (PFN)~\cite{Komiske:2018cqr}, an application of the deep sets theorem~\cite{Zaheer:2017}, that uses per-particle information of the particles associated with the jet as well as per-event information associated with the nonprompt lepton, the jet, and the SV.
The PFN architecture consists of two sections, with three fully connected neural network layers of 128 nodes each per section.
First, each individual particle is passed through the per-particle section, which is provided with the per-particle information and produces a 256-dimensional output representing the latent space.
The per-particle outputs are added up in a summation layer, and the sums are provided together with event-level variables as inputs to the second section that processes the full event information.
The two sections are combined into a single network, which allows for a simultaneous optimization of both the per-particle and per-event sections.
The PFN output is a single score value between zero and one, where high (low) values are assigned to signal-like (background-like) events.

Up to 50 PF particles associated with the selected jet are used for the per-particle section.
For each particle, the \pt, \sigeta, $\phi$, charge, particle type as identified by the PF algorithm, impact parameter properties and uncertainties, number of hits in the pixel and strip detectors of the tracker system~\cite{CMS:TRK-11-001}, and the association with tracks and SVs are provided as input variables to the PFN.

The per-event input variables are associated with the selected jet, nonprompt lepton, and SV.
For the jet, the kinematic properties and the number of constituent particles are provided.
For the nonprompt lepton, the kinematic properties, \Irel, impact parameter properties and uncertainties, and the number of hits in the pixel detector are used.
Additionally, the \dR between lepton and jet, the transverse component of the lepton momentum with respect to the jet axis, and the \pt ratio between the lepton and the jet are provided.
For the SV, the kinematic properties including the invariant mass \mSV of the tracks associated with the SV, the number of tracks, the normalized $\chi^2$ of the reconstruction as determined with the IVF, and the distance from the PV are used.

The PFN is trained with MC samples for the HNL signal, and the SM background samples described in Section~\ref{sec:samples}.
Only the QCD multijet sample is excluded from the training, because of its large statistical uncertainties.
Only events passing the event selection summarized in Table~\ref{tab:eventselection} are used for the training.
To account for the upgrade of the pixel detector in 2017~\cite{CMS:Pixel-2021}, PFNs are trained separately for the 2016 data set and the combined 2017--2018 data sets.
Separate PFNs are trained for nonprompt electrons and muons, and for \mhnl up to and above 5\GeV.
The reason for the split into a low- and high-mass PFN is the strong dependence of some of the input variables for HNL signal events on the \mhnl scenario, and also heavy-flavor background events are more similar to low-mass than to high-mass HNL events.
For the low-mass (high-mass) PFN, the training is performed using an admixture of MC signal samples with HNL masses in the range of 1--5 (6--20)\GeV and with different \tauhnl values.

\begin{figure}[!ht]
\centering
\includegraphics[width=0.42\textwidth]{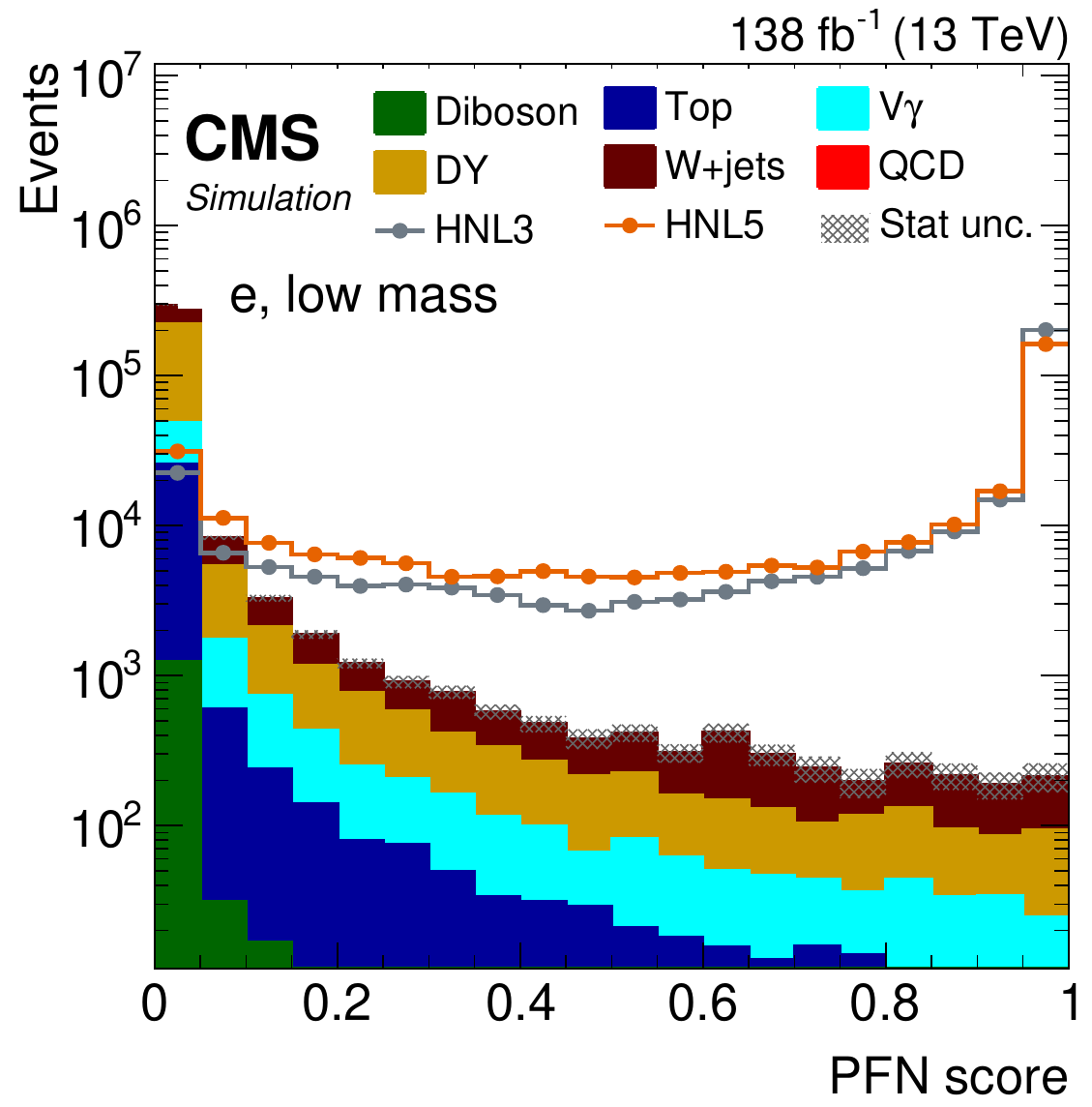}%
\hspace{0.05\textwidth}%
\includegraphics[width=0.42\textwidth]{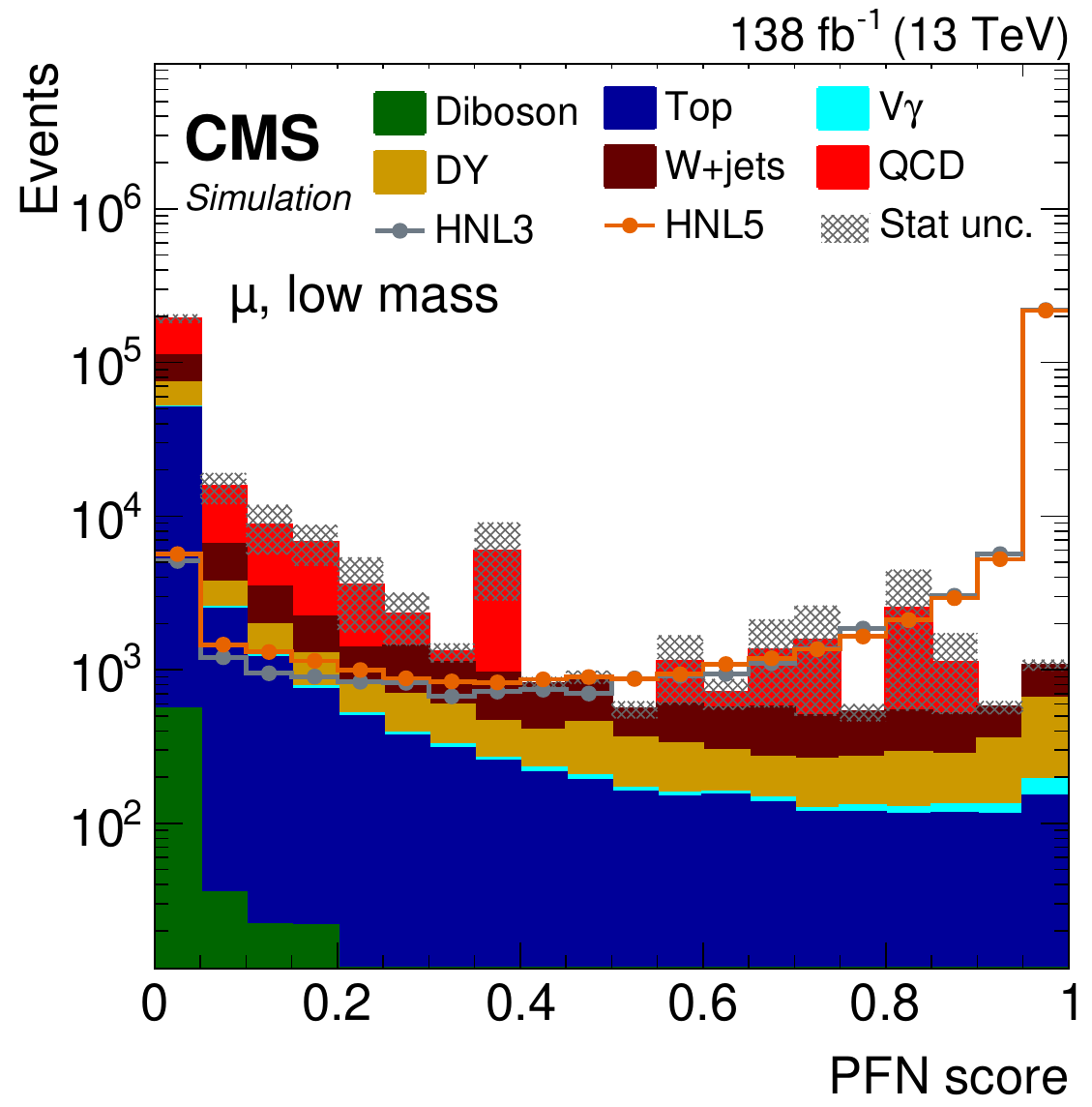} \\[1ex]
\includegraphics[width=0.42\textwidth]{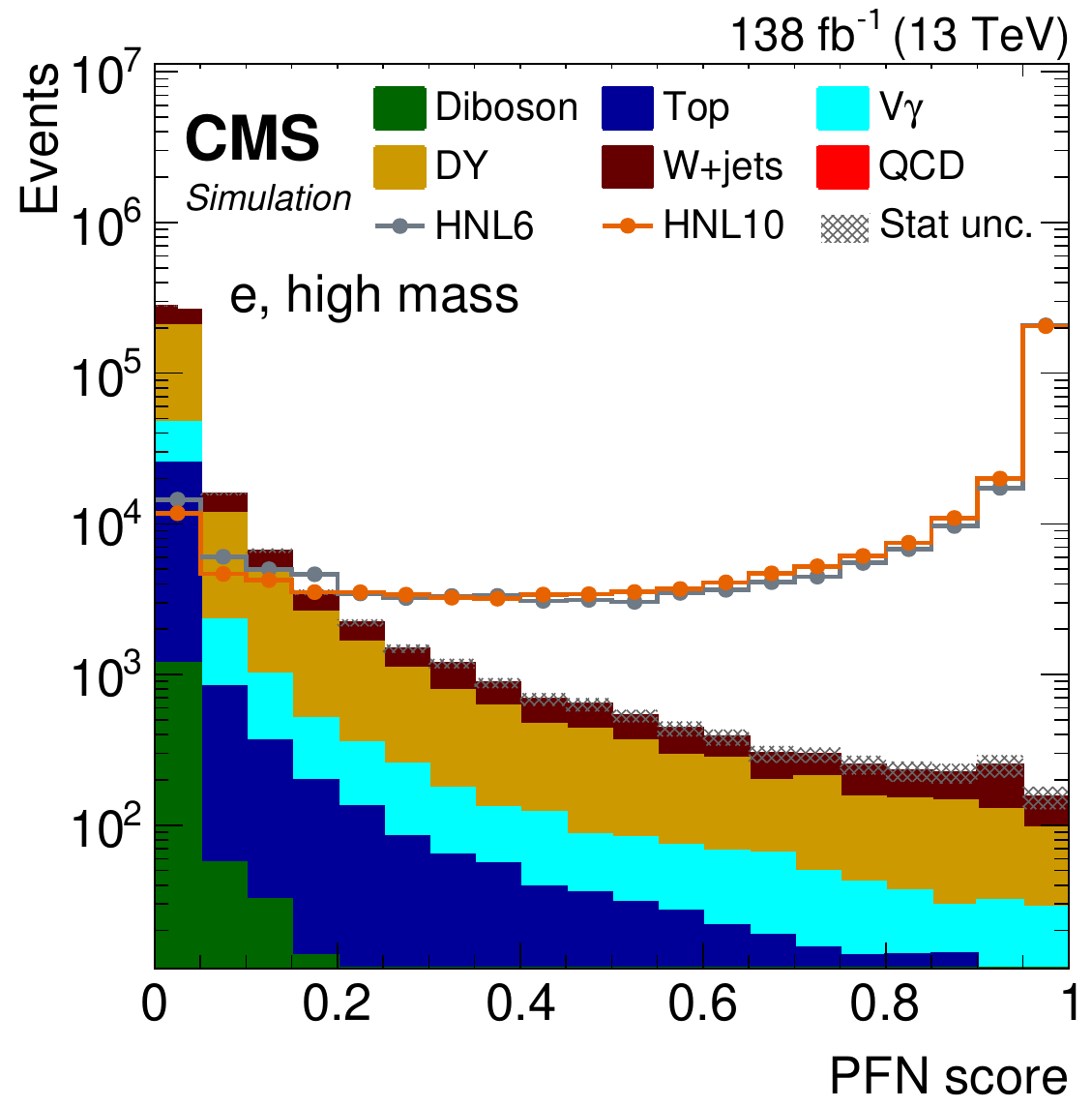}%
\hspace{0.05\textwidth}%
\includegraphics[width=0.42\textwidth]{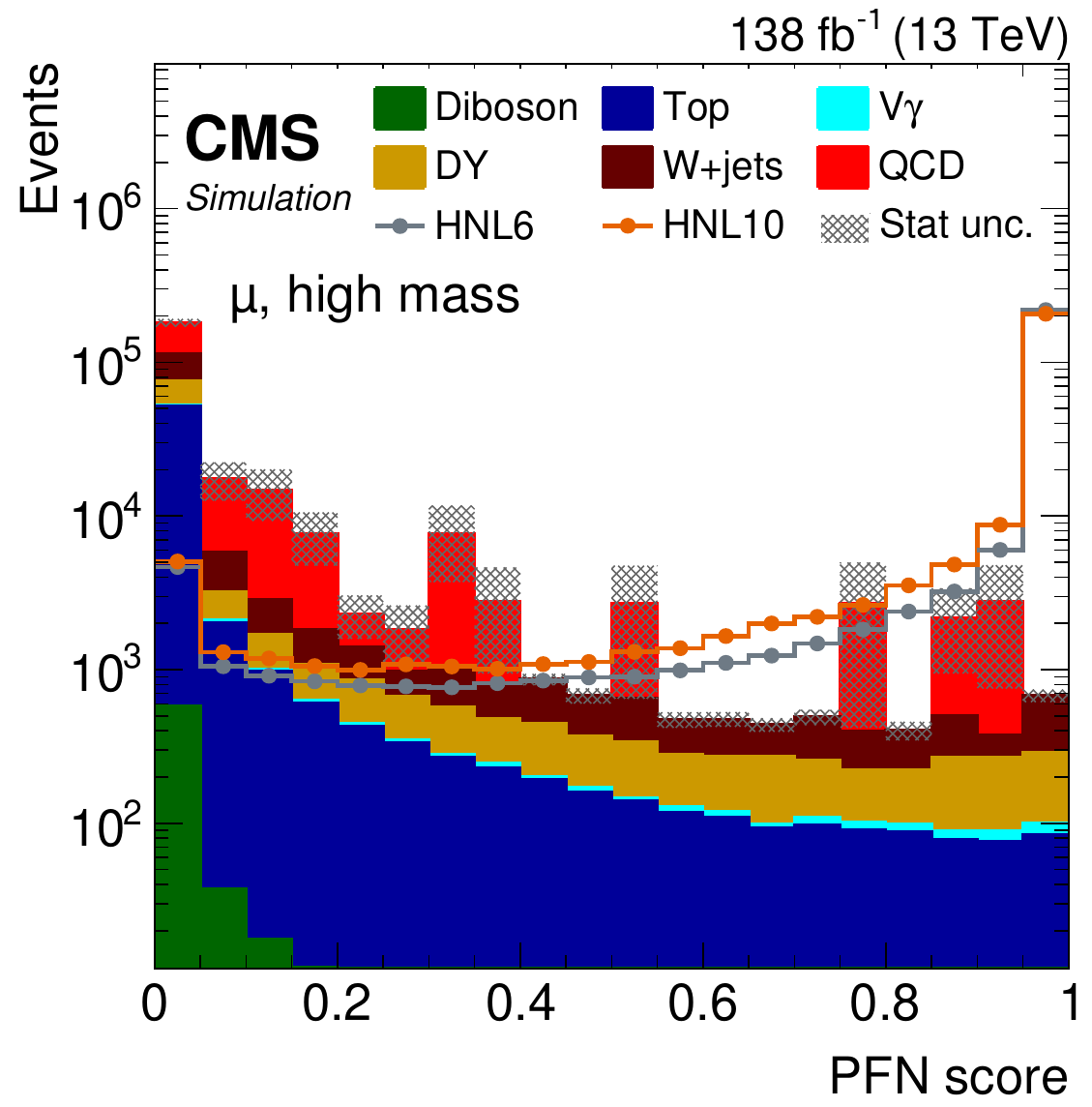}
\caption{%
    The PFN score distribution of predicted events yields after applying the selection summarized in Table~\ref{tab:objectselection}, for the combined \Cee and \Cme (left) or \Cmm and \Cem (right) categories, using the low-mass (upper) or high-mass (lower) PFNs.
    The filled histograms display the predicted background yields, where the QCD multijet background displays huge statistical fluctuations for PFN scores above 0.2.
    The lines show the predicted yields for four HNL signal scenarios, with $\mhnl=3\GeV$ and $\mixparsq=9.9\times10^{-5}$ (HNL3), $\mhnl=5\GeV$ and $\mixparsq=1.6\times10^{-6}$ (HNL5), $\mhnl=6\GeV$ and $\mixparsq=2.0\times10^{-6}$ (HNL6), and $\mhnl=10\GeV$ and $\mixparsq=5.7\times10^{-7}$ (HNL10).
    The HNL signal yield is normalized to the total background yield.
}
\label{fig:pfnscore}
\end{figure}

The PFN training is performed with the \textsc{keras} v2.1.5 deep learning library~\cite{Chollet:2015} interfaced to \textsc{TensorFlow} v1.6~\cite{Abadi:2015}.
Parametric rectified linear functions are chosen for the activation function.
The dropout parameter is set to a high value of 50\%, \ie, a random half of the neurons are dropped from each training iteration to prevent overtraining~\cite{Srivastava:2014jmlr}.
We set aside 10\% of the training samples as a testing set, which is not used during the training but only for the final validation after the full training is done.
Two thirds of the remaining events in the training samples are used for training during an epoch and the other third for validation after the epoch.
The PFN is trained for a maximum of 100 epochs, and stopped earlier if the accuracy on the training data does not increase for four full training epochs.
This cycle is repeated three times, where a different third of the events is used as the validation set.
This $K$-fold cross-validation method, where each event is used both for training and validation, further reduces the possibility of overtraining~\cite{Hastie:2009spr}.
The final model is obtained by taking the average of the three models with the best accuracy on the validation data of the respective training cycle.
The PFN score is shown for simulated HNL and background events in Fig.~\ref{fig:pfnscore}.

\begin{figure}[!ht]
\centering
\includegraphics[width=0.45\textwidth]{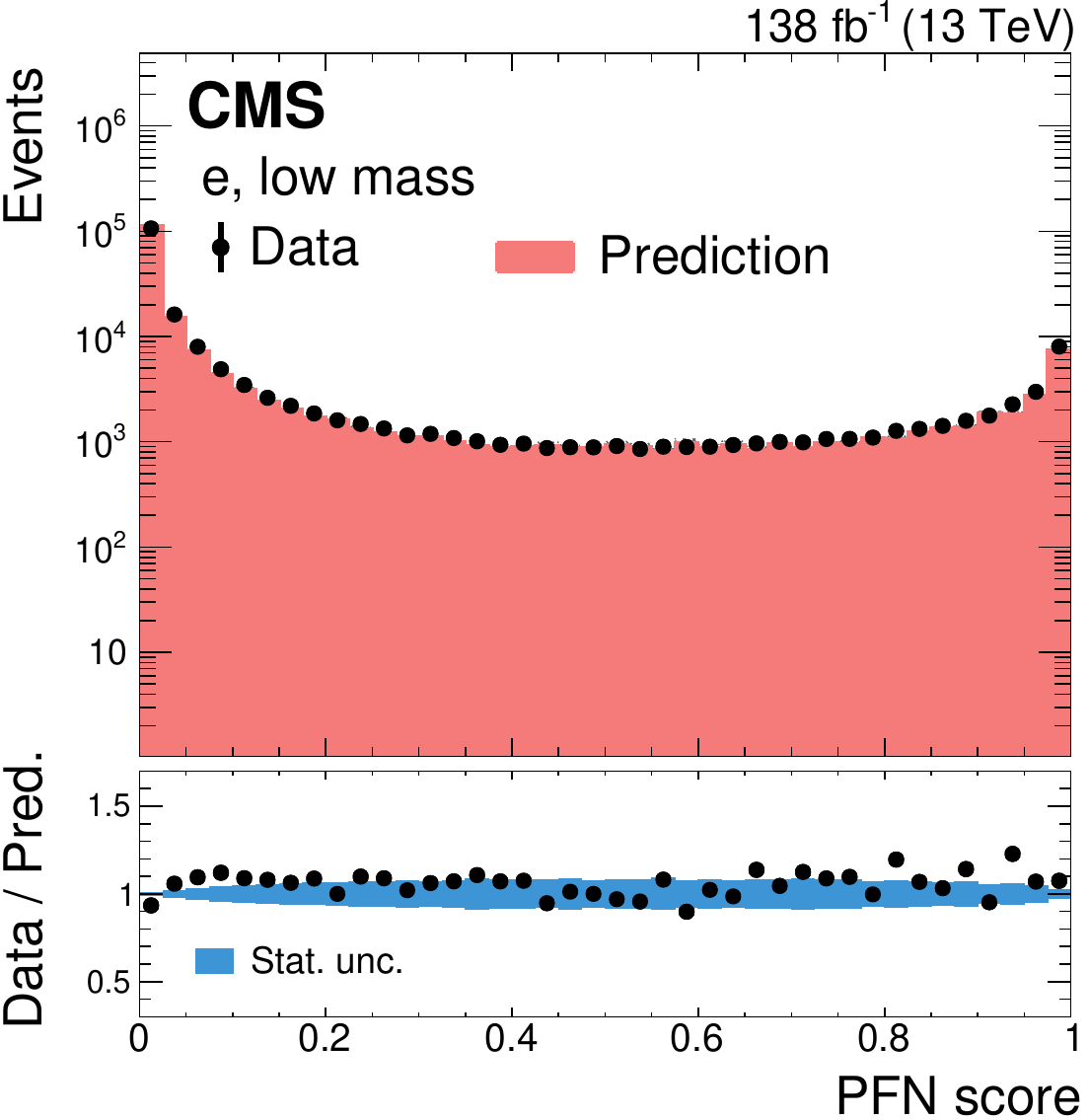}%
\hspace{0.05\textwidth}%
\includegraphics[width=0.45\textwidth]{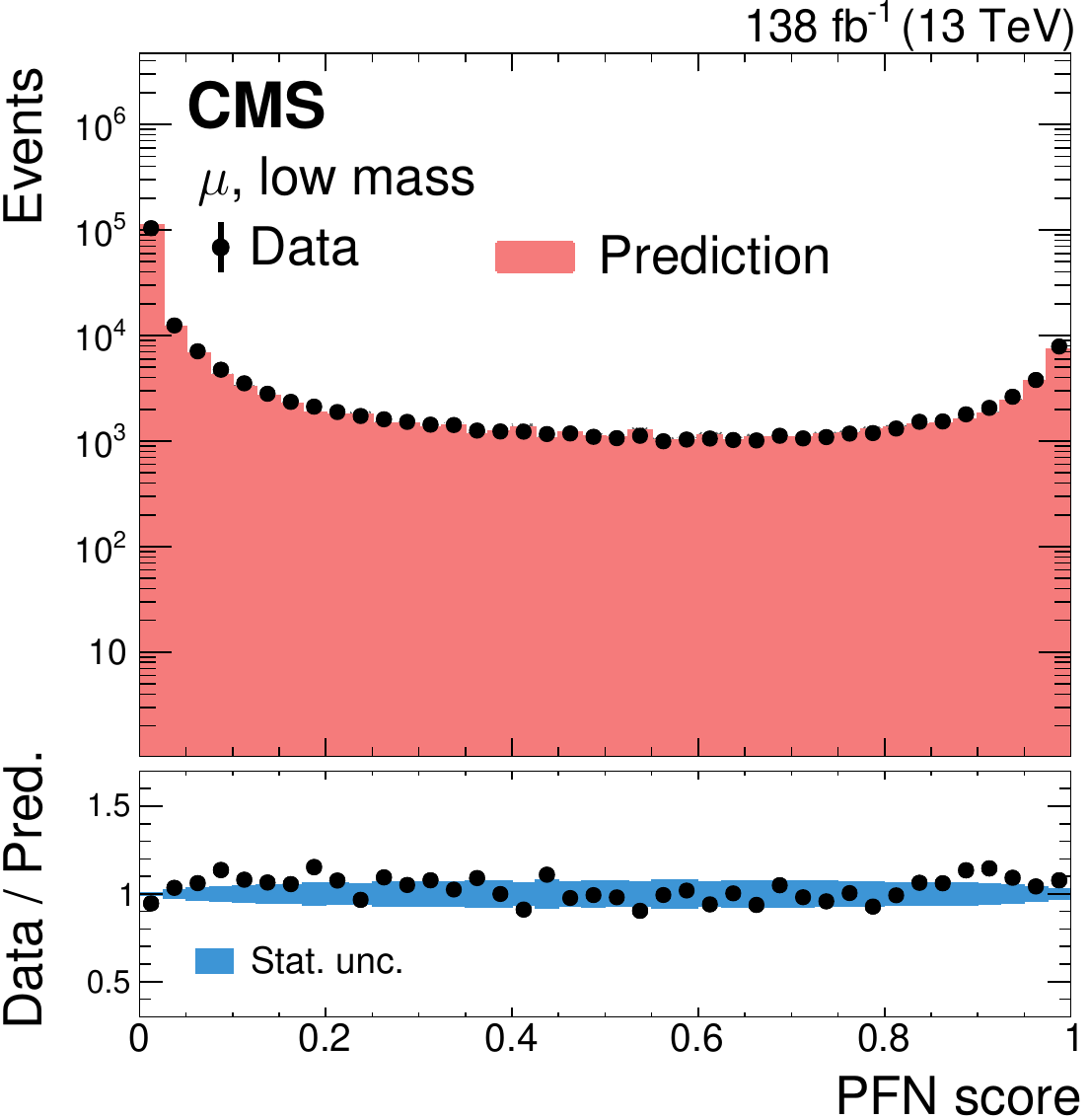}%
\caption{%
    Predicted and observed event yields in the PFN score distribution for $\PKzS\to\PGpp\PGpm$ decays in $\PZ\to\MM$ events, using the low-mass PFNs for the electron (left) and muon (right) channels, where the \PGppm with higher \pt is treated as the lepton.
    The prediction is scaled to match the overall data yield.
    The lower panels show the data-to-prediction ratio.
}
\label{fig:kshort}
\end{figure}

A validation of differences in the shape of the PFN score between data and simulated events is performed to ensure that the signal efficiency is the same in data and simulation.
The meson decays $\PKzS\to\PGpp\PGpm$ are a representative long-lived decay process that can be used to emulate HNL decays.
Following a procedure similar to that described in Ref.~\cite{CMS:DP-2024-010}, a pure sample of $\PZ\to\MM$ events is selected, requiring two OS muons with \mmpmmp within 10\GeV of the \PZ boson mass~\cite{ParticleDataGroup:2022pth} to facilitate the data-to-simulation comparison and normalization.
These events are then required to contain an SV formed by two tracks with an invariant mass within 15\MeV of the \PKzS meson mass~\cite{ParticleDataGroup:2022pth}.
The \PKzS meson is typically found within a jet with additional PF candidates.
The identification of the SV and jet is performed with the same configuration employed for HNL events.
All PF particles in the jet, including those not associated with the \PKzS meson decay, are used as inputs to the PFN.
Since no displaced lepton is present in the \PKzS meson decay, the \PGppm candidate track with the higher \pt is chosen to emulate the displaced lepton for the PFN.
In Fig.~\ref{fig:kshort}, the result of this comparison is shown for the low-mass PFNs.
The total \PKzS candidate yield derived from MC samples is normalized to the yield in data, to facilitate the shape comparison of the PFN score.
Agreement between the data and the MC prediction within 20\% is found over the whole range of the PFN score, both in the electron and muon channels, with an agreement of better than 10\% for the last two bins with the highest PFN scores.
A similar level of agreement is found for the high-mass PFNs.
This validation is used to assign a systematic uncertainty for data-to-simulation differences in the PFN score, as discussed in Section~\ref{sec:systematics}.

\section{Background estimation}
\label{sec:background}

After applying all event selection criteria, the main sources of SM background contributions are top quark, \Wjets, DY, and QCD multijet production.
Their relative contributions vary across the different channels.
To further suppress background events and enrich the event sample used for the background estimation with events that have properties similar to expected HNL signal events, all events are required to have $\mlSV>10\GeV$ and a PFN score of at least 0.2.
The selection after these additional requirements will be referred to as the signal region (SR), and is generally different when using low- or high-mass PFNs.
Contributions from DY production are found in the \Cee and \Cmm categories, where one lepton is poorly reconstructed and identified as nonprompt, and combined with a random track to form a compatible SV.
Background contributions from top quark (\Wjets) production typically have one genuine prompt lepton and a second lepton inside a jet that is reconstructed in genuine (accidental) association with an SV.
The QCD multijet events contribute especially to the \Cmm and \Cem categories, and neither the prompt nor nonprompt lepton is genuine.
A final background source arises from displaced photon conversions in the detector material and contributes mainly to the \Cee and \Cme categories, which have a photon radiated from a prompt lepton and subsequently converted into an $\EE$ pair with one electron receiving most of the momentum.

\begin{figure}[!ht]
\centering
\includegraphics[width=0.45\textwidth]{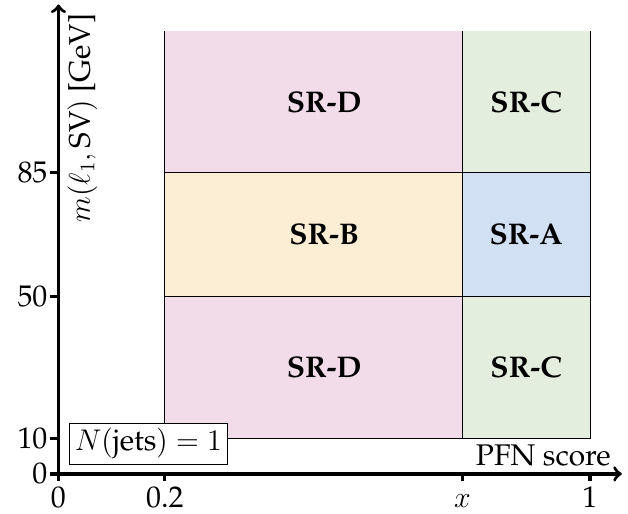}%
\caption{%
    Illustration of the target and sideband region definitions for the ABCD method applied to the SR, in terms of \Njets, \mlSV, and the PFN score.
}
\label{fig:abcd}
\end{figure}

Background estimates for displaced signatures based on MC simulation are known to be poorly modeled.
Thus, we employ a background estimation method using data control samples and rely on the discriminating powers of the \mlSV distribution described in Section~\ref{sec:eventselection} and the PFN score described in Section~\ref{sec:PFN}.
A target region A is defined by requiring $50<\mlSV<85\GeV$ and that the PFN score is above a threshold value \PFNthreshold.
An ABCD method~\cite{Choi:2019mip} is then employed to predict the background contribution in region A from the observed yields in three sideband regions B, C, and D, based on inverting one or both of the \mlSV and PFN score requirements.
The region definitions are listed in Table~\ref{tab:abcd} and shown schematically in Fig.~\ref{fig:abcd}.
The expected number of background events in region A, \ABCD{A}, is then estimated from the observed number of events in the sideband regions as $\ABCD{A}=\ABCD{B}\ABCD{C}/\ABCD{D}$.

\begin{table}[!ht]
\centering
\topcaption{%
    Definition of target and sideband regions used in the ABCD background estimation method for the signal (SR), validation (VR), and control (CR) regions.
    The threshold value $x$ is chosen between 0.97 and 0.998 separately for each event category, as described in the text.
}
\label{tab:abcd}
\renewcommand\arraystretch{1.2}
\begin{tabular}{cccc}
    Region & \Njets & PFN score & \mlSV [{\GeVns}] \\ \hline
    SR-A & $=$1 & $>$\PFNthreshold & $>$50, $<$85 \\
    SR-B & $=$1 & $>$0.2, $<$\PFNthreshold & $>$50, $<$85 \\
    SR-C & $=$1 & $>$\PFNthreshold & $>$10, $<$50 or $>$85 \\
    SR-D & $=$1 & $>$0.2, $<$\PFNthreshold & $>$10, $<$50 or $>$85 \\[\cmsTabSkip]
    VR-AB & $>$1 & $>$0.2 & $>$50, $<$85 \\
    VR-CD & $>$1 & $>$0.2 & $>$10, $<$50 or $>$85 \\[\cmsTabSkip]
    CR-A & $\geq$1 & $>$\PFNthreshold & $>$85, $<$110 \\
    CR-B & $\geq$1 & $>$0.2, $<$\PFNthreshold & $>$85, $<$110 \\
    CR-C & $\geq$1 & $>$\PFNthreshold & $>$10, $<$50 or $>$110 \\
    CR-D & $\geq$1 & $>$0.2, $<$\PFNthreshold & $>$10, $<$50 or $>$110 \\
\end{tabular}
\end{table}

To optimize the values \PFNthreshold of the PFN threshold, we consider the signal-over-background ratio in region A as well as the statistical precision of the ABCD background estimate.
For the latter, it is important that \ABCD{C} does not become zero, which would result in an estimate of $\ABCD{A}=0$ with a large statistical uncertainty.
The result of this optimization are values of $0.97\leq\PFNthreshold\leq0.998$, chosen separately for each event category.
To further increase the sensitivity to the HNL signal, the SR is binned in \mSV and the transverse displacement \Deltwod between PV and SV.
The ABCD background estimation is done separately for each bin in \mSV and \Deltwod, and separately using the low- and high-mass PFNs.

\begin{figure}[!ht]
\centering
\includegraphics[width=0.45\textwidth]{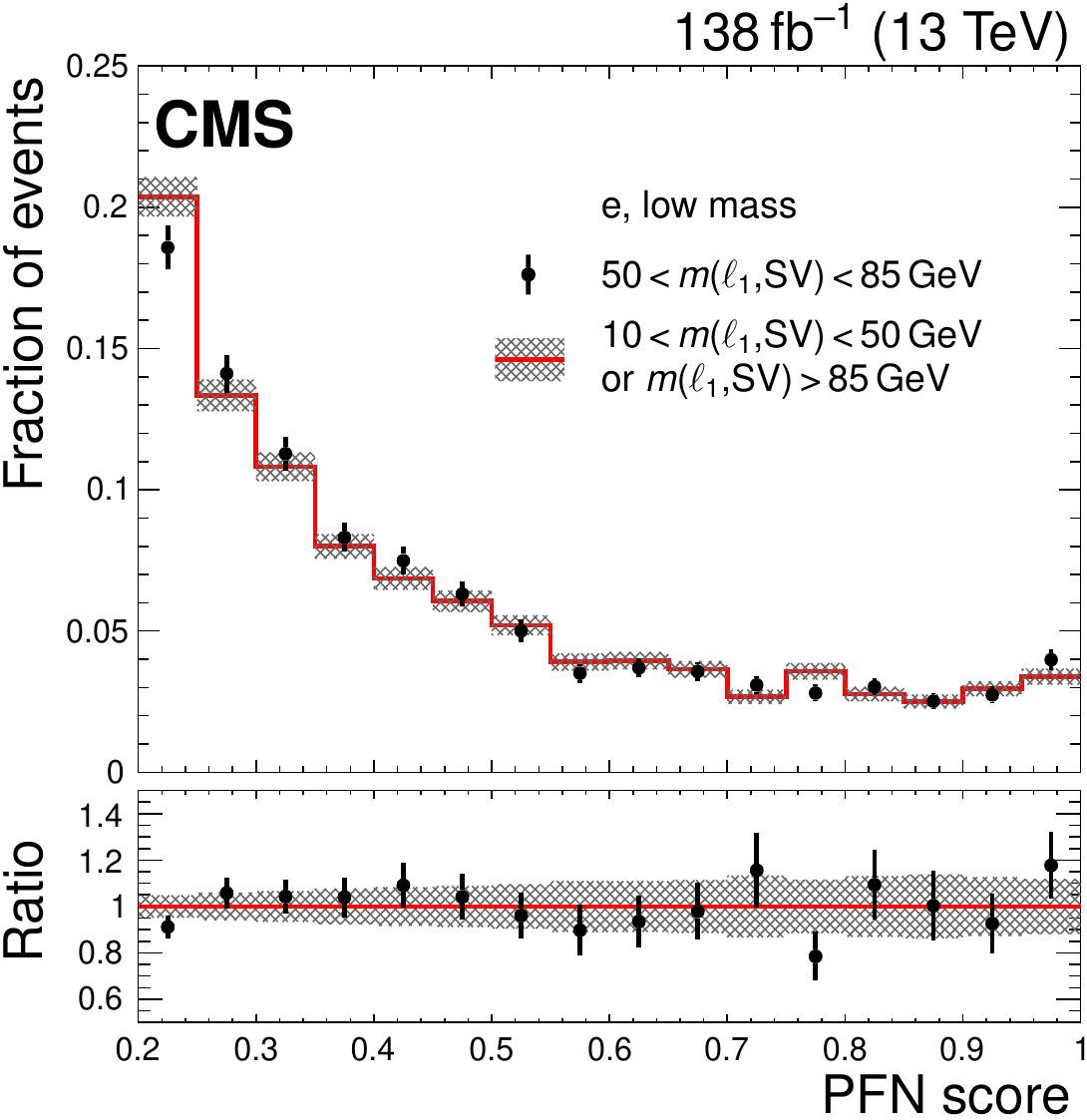}%
\hspace{0.05\textwidth}%
\includegraphics[width=0.45\textwidth]{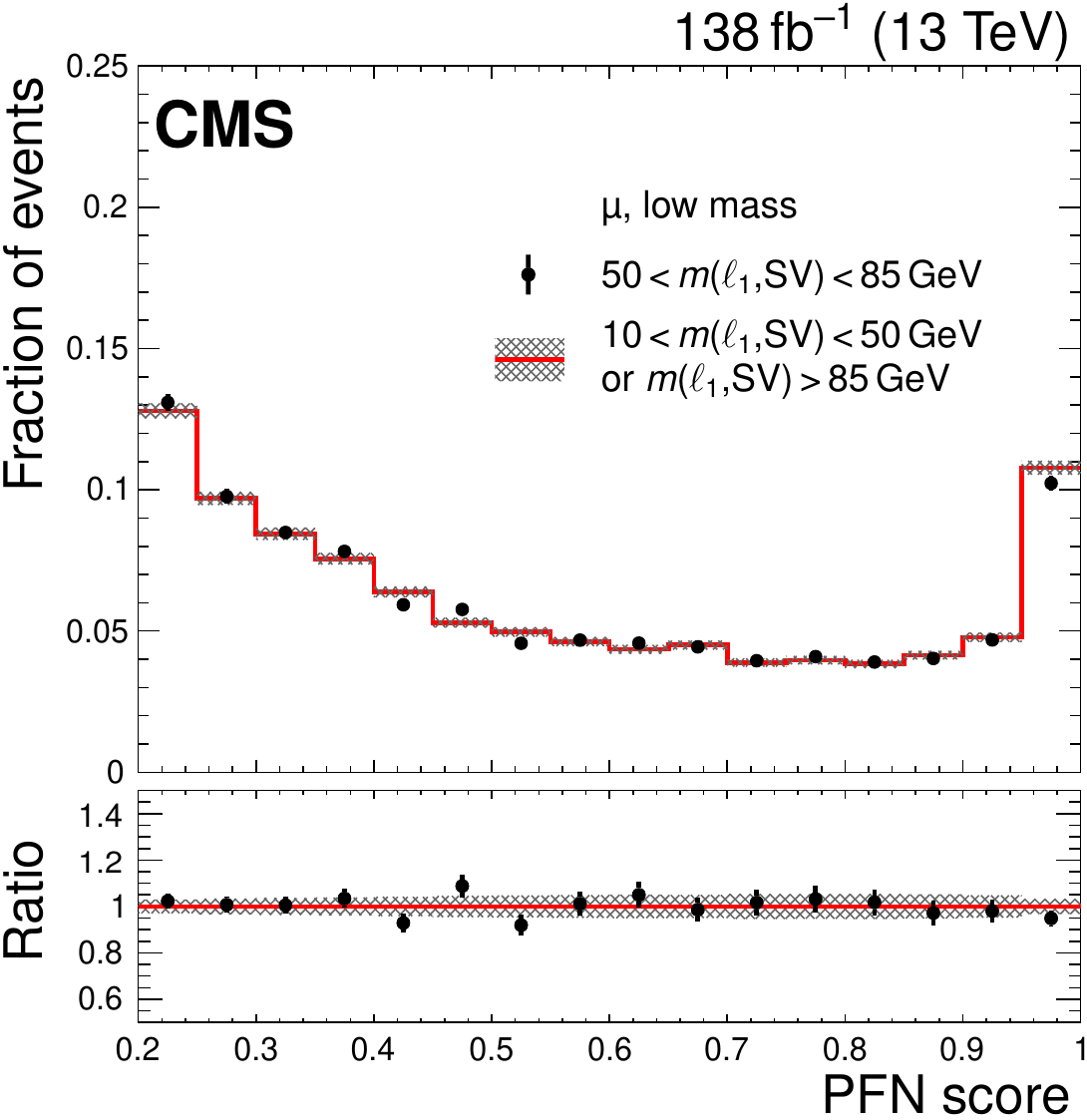}%
\\[1ex]
\includegraphics[width=0.45\textwidth]{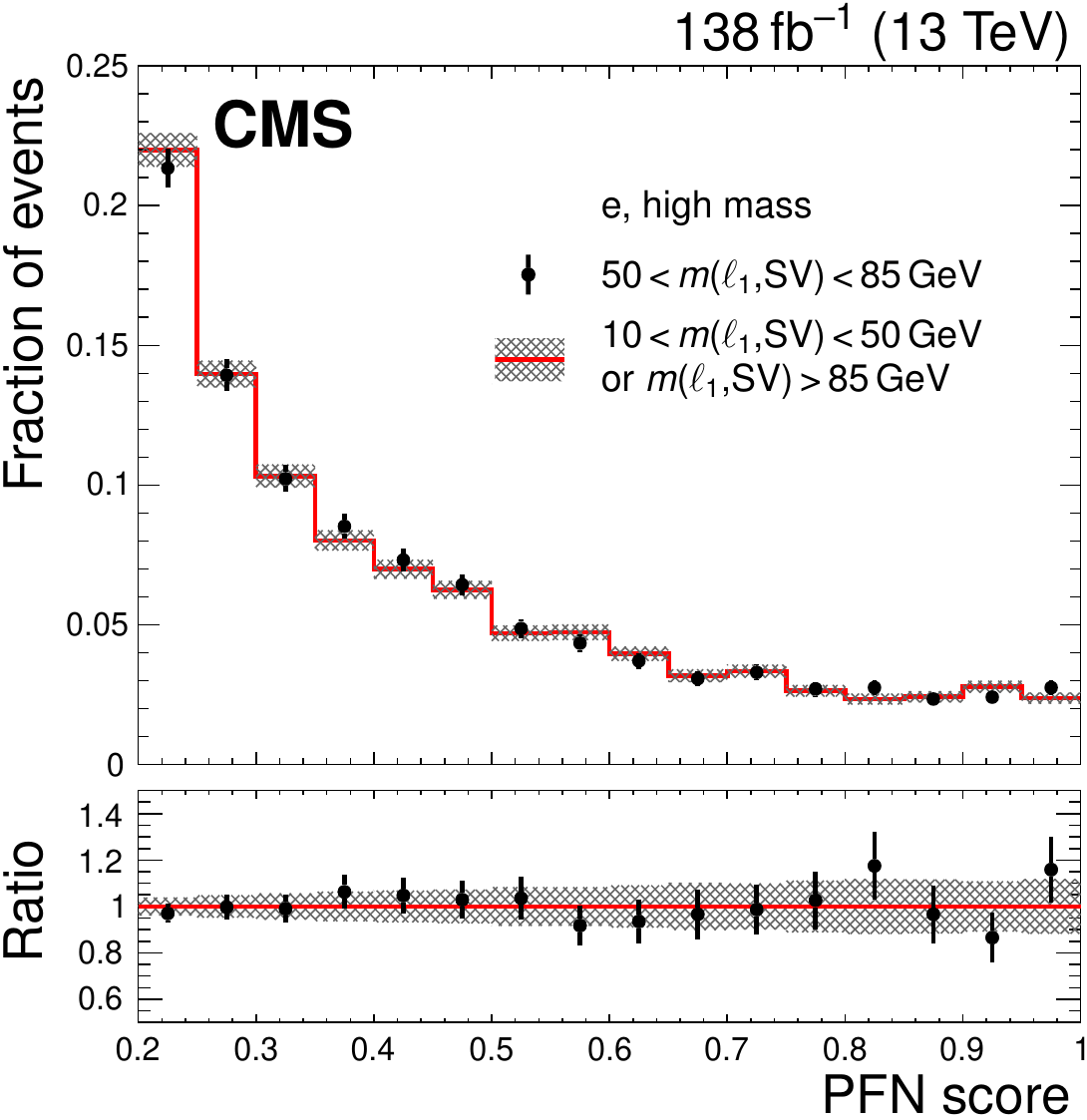}%
\hspace{0.05\textwidth}%
\includegraphics[width=0.45\textwidth]{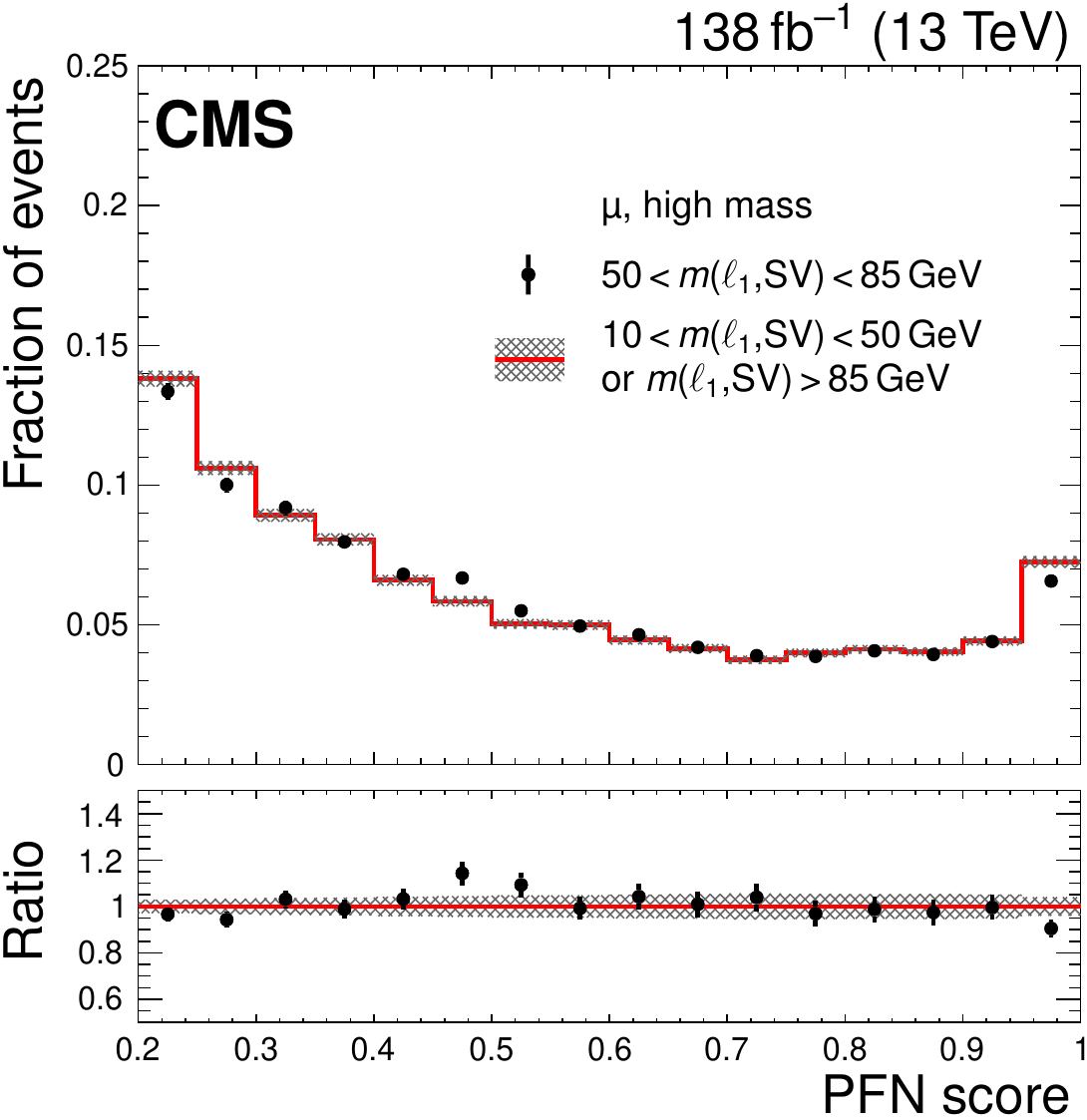}%
\caption{%
    The PFN scores in data in the VR, shown for the electron (left) and muon (right) trainings with the low-mass (upper) and high-mass (lower) samples.
    The distributions are normalized to unity to facilitate a shape comparison.
    Statistical uncertainties are indicated with error bars and shaded areas.
    The lower panels show the ratio of the VR-AB to the VR-CD yields.
}
\label{fig:vr}
\end{figure}

The ABCD method relies on the absence of correlation between the PFN score and \mlSV.
To evaluate in data whether this assumption is valid, we form an orthogonal event selection with $\Njets>1$, referred to as validation region (VR).
For the first test, we divide the VR into a target region with $50<\mlSV<85\GeV$ and a sideband region outside of this interval, as listed in Table~\ref{tab:abcd}.
In Fig.~\ref{fig:vr}, the PFN scores in data are compared for VR events in the target and sideband regions and, notably, no difference is observed between the shapes.

As further tests, we calculate Pearson correlation coefficients, which measure the strength and direction of the linear relation, to quantify the correlation between \mlSV and the PFN score, as well as $p$-values to determine the statistical significance of the correlation~\cite{Remillard:2011iess}.
This is done separately in flavor channels and $(\mSV,\Deltwod)$ bins, but inclusively in the ABCD classification of the VR.
We find correlation coefficients very close to zero in most cases, and also the $p$-values indicate a weak or negligible linear dependence of the PFN score on \mlSV.
Only in the OS \Cmm category of the VR with the low-mass PFN, a relevant statistical dependence is found.
We identified in studies with simulated event samples that DY dimuon production is the only background process that exhibits this correlation.

To correct for the bias in the background estimation from the correlation between \mlSV and the PFN score in DY dimuon events, we apply correction factors to the ABCD estimate.
The correction factor $c_{\text{DY}}$ that needs to be applied to the DY contribution in a bin is calculated as the ratio between the direct DY MC prediction for \ABCD{A} in that bin with the ABCD estimate $\ABCD{B}\ABCD{C}/\ABCD{D}$ evaluated from the same DY MC prediction.
Using the fraction $f_{\text{DY}}$ of DY MC events relative to the total MC background prediction in that bin, the correction factor to be applied to the ABCD background prediction evaluated from data is $c_{\text{bkg}}=1+(c_{\text{DY}}-1)f_{\text{DY}}$.
Each bin of the OS \Cmm categories when using the low-mass PFNs is corrected separately.
The limited size in the DY MC event samples used for the evaluation of $c_{\text{bkg}}$ results in a systematic uncertainty, as discussed in Section~\ref{sec:systematics}.

\begin{figure}[!p]
\centering
\includegraphics[width=0.95\textwidth]{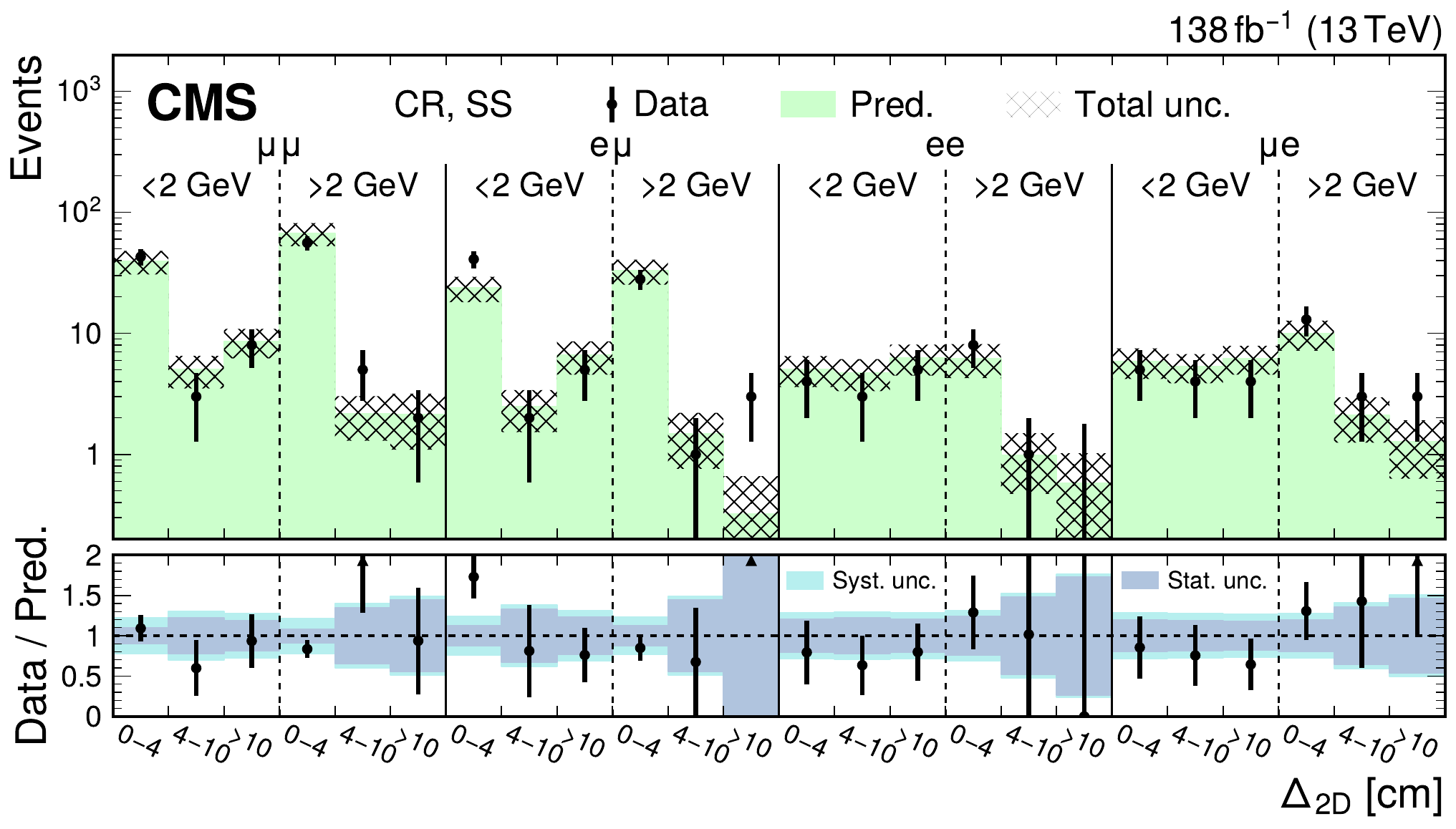}%
\\[1ex]
\includegraphics[width=0.95\textwidth]{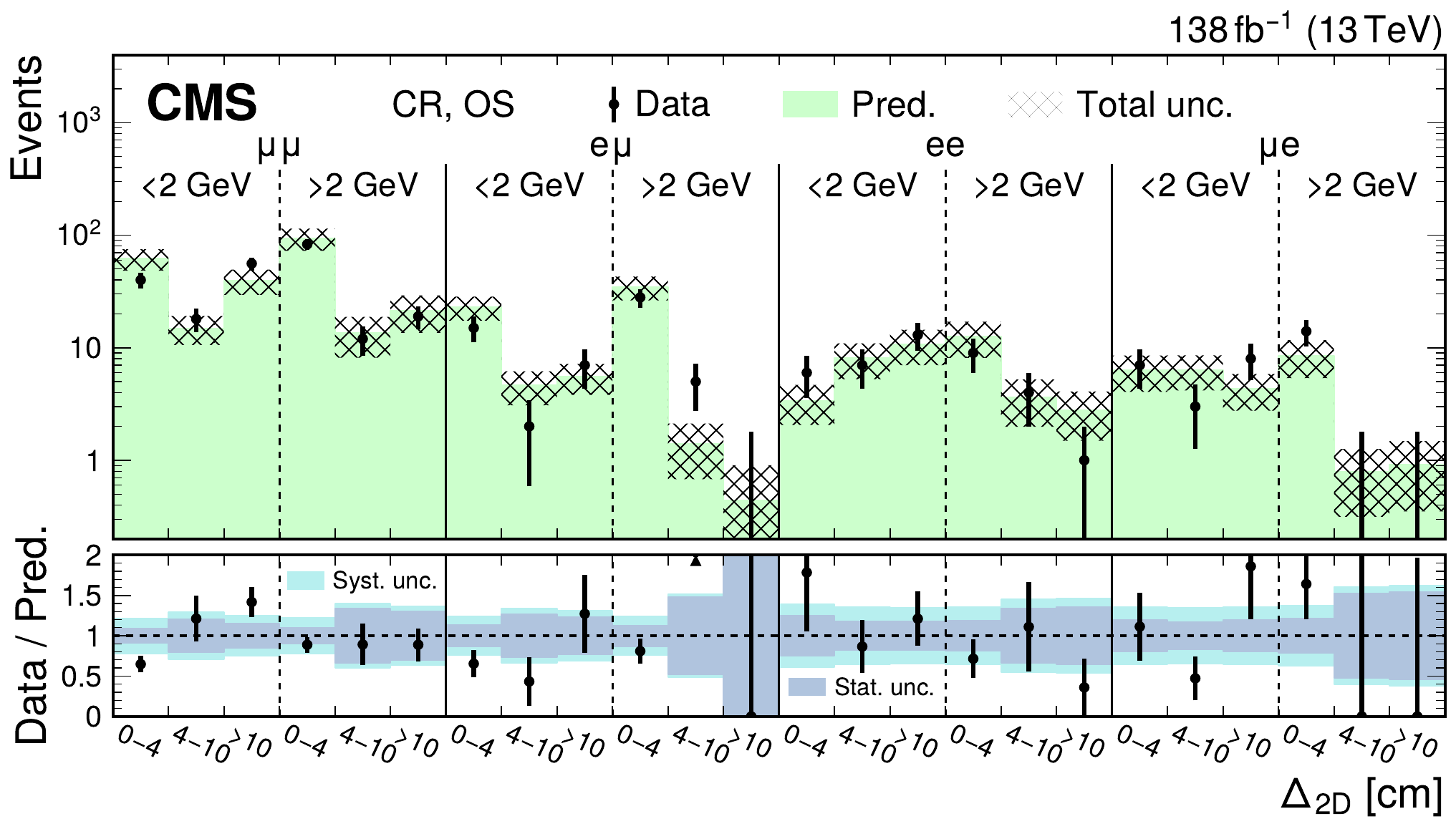}%
\caption{%
    Predicted and observed event yields in the CR for the low-mass PFNs in the SS (upper) and OS (lower) categories, binned by flavor channel, \mSV (as specified below the flavor channel), and \Deltwod.
    The hashed bands represent the total uncertainty in the background prediction.
    The lower panels show the data-to-prediction ratio and the background prediction uncertainty is split into statistical (``stat.'') and systematic (``syst.'') contributions.
}
\label{fig:crlow}
\end{figure}

\begin{figure}[!p]
\centering
\includegraphics[width=0.95\textwidth]{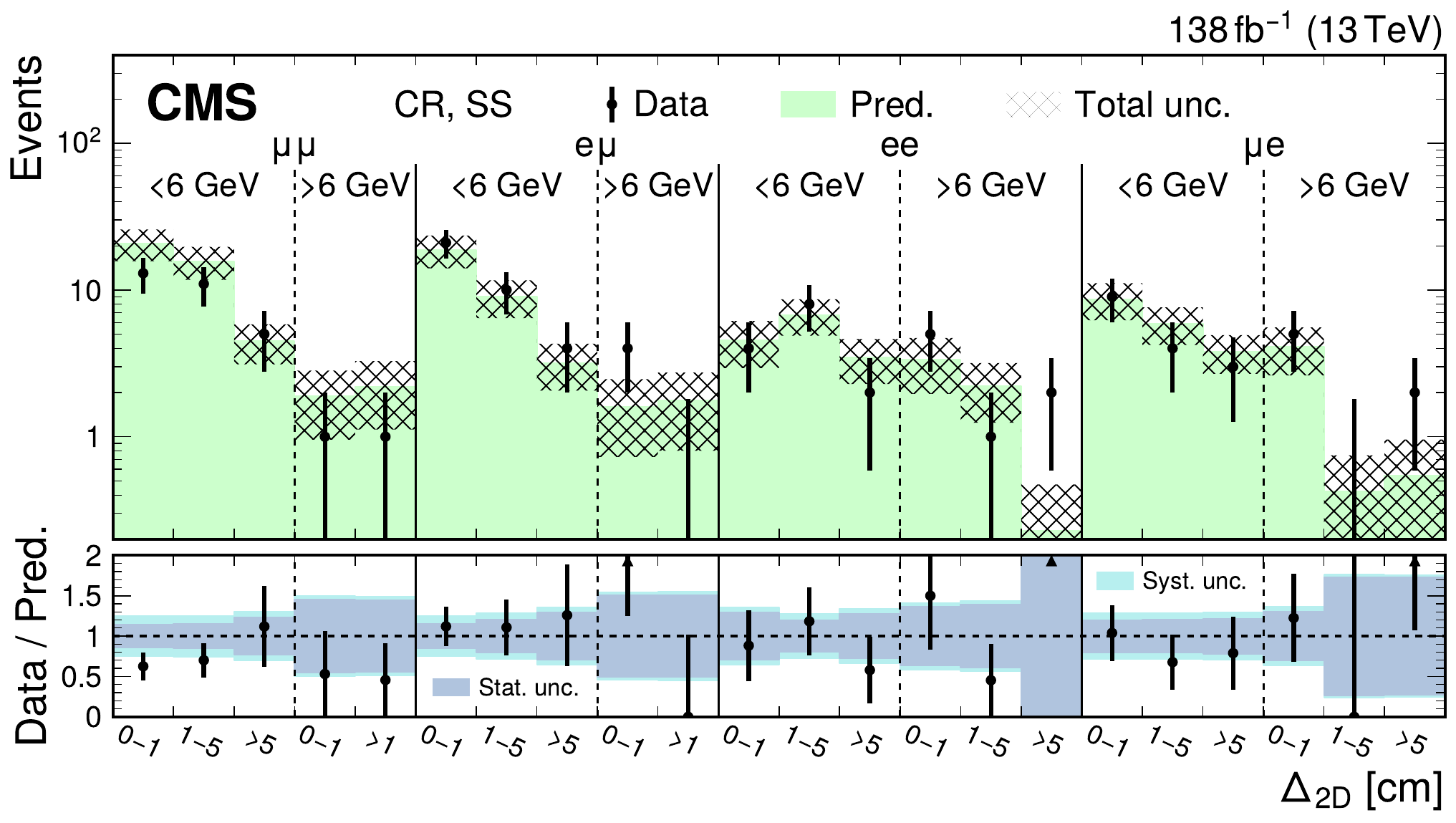}%
\\[1ex]
\includegraphics[width=0.95\textwidth]{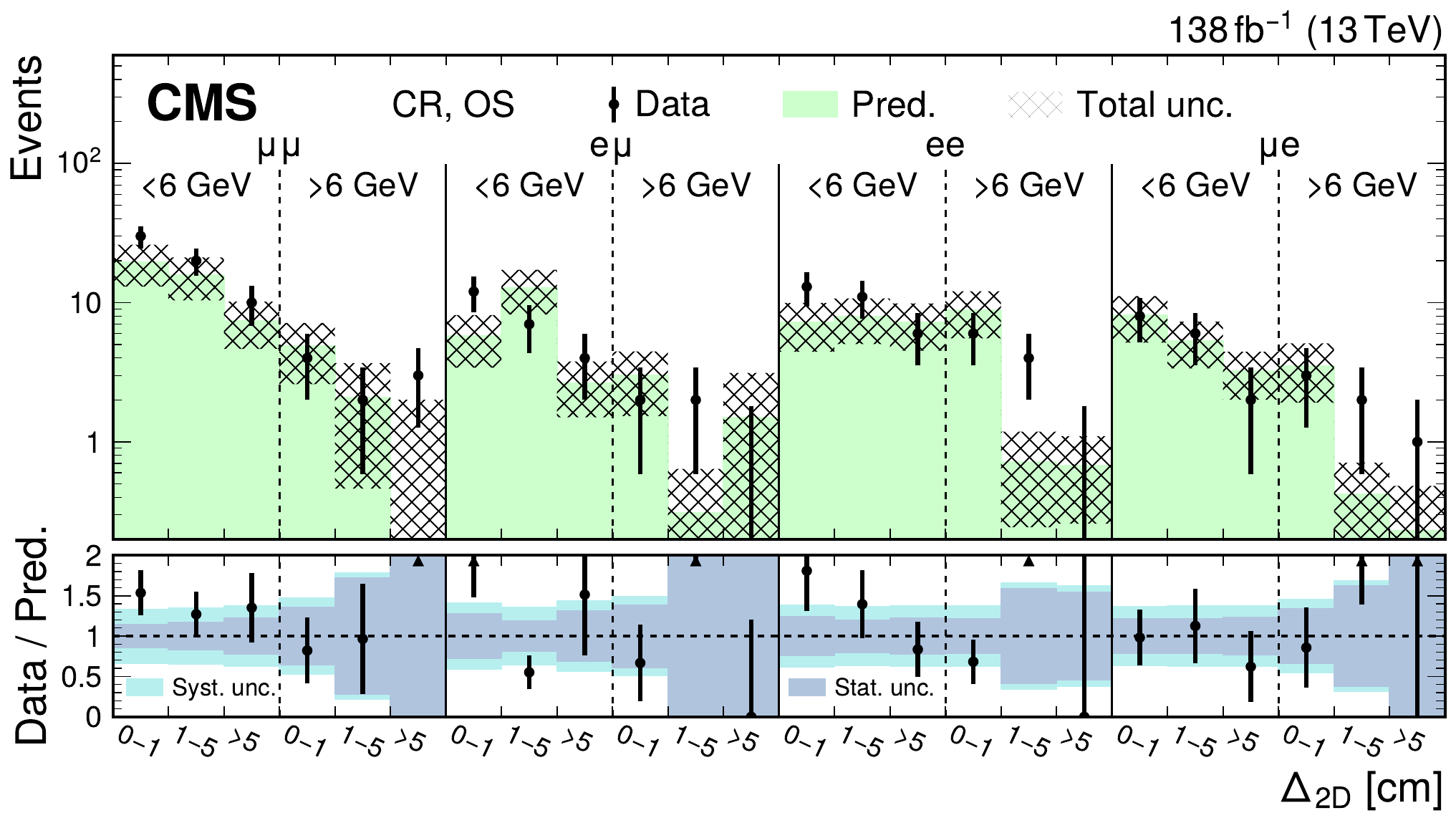}%
\caption{%
    Predicted and observed event yields in the CR for the high-mass PFNs in the SS (upper) and OS (lower) categories, binned by flavor channel, \mSV (as specified below the flavor channel), and \Deltwod.
    The hashed bands represent the total uncertainty in the background prediction.
    The lower panels show the data-to-prediction ratio and the background prediction uncertainty is split into statistical (``stat.'') and systematic (``syst.'') contributions.
}
\label{fig:crhigh}
\end{figure}

We evaluate the closure of the ABCD background estimation in data by defining a control region (CR) that does not include the target region SR-A, splitting it into target and sideband regions, and comparing the ABCD background estimate for CR-A with the observed yields.
The CR is formed from all events in the SR and VR outside the interval $50<\mlSV<85\GeV$, \ie, excluding the target regions of the SR and VR.
The inclusion of the VR increases the statistical power of this closure test.
The target region of the CR is defined by the requirement $85<\mlSV<110\GeV$, and the corresponding sideband definitions are listed in Table~\ref{tab:abcd}.
The comparison between the ABCD estimate for CR-A with the observed yield is done in bins of \mSV and \Deltwod, optimized separately for usage with the low- and high-mass PFNs to provide discrimination between signal and background, as well as between different signal hypotheses.
For the low-mass analysis, two \mSV bins $<$2 and $>$2\GeV are formed, and each is divided into three \Deltwod bins $<$4, 4--10, and $>$10\unit{cm}.
For the high-mass analysis, the two \mSV bins are $<$6 and $>$6\GeV, each divided by \Deltwod as $<$1, 1--5, and $>$5\unit{cm}.
For the SS \Cmm and \Cem channels with the high-mass PFNs, we split the $\mSV>6\GeV$ bin into only two \Deltwod bins with $<$1 and $>$1\unit{cm}, to avoid having a $\Deltwod>5\unit{cm}$ bin with zero predicted background.
The results of the comparison are shown in Figs.~\ref{fig:crlow}--\ref{fig:crhigh}.
A good closure of the background prediction estimate is found within 20--30\% when taking the large statistical uncertainty of the data into account.
As discussed in Section~\ref{sec:systematics}, a systematic uncertainty in the background prediction is assigned based on this closure.

\section{Systematic uncertainties}
\label{sec:systematics}

The predicted number of background events, as well as the expected number of signal events in the different SR bins for the different event categories, are affected by several sources of systematic uncertainty.
Several theoretical and experimental uncertainties affect the HNL signal yields, with some changing only the overall normalization and others also affecting the shape of the signal prediction in the SR bins.
For the background yields, two dedicated uncertainties are evaluated that both affect the shape of the prediction.
A summary of the different systematic uncertainty sources is given in Table~\ref{tab:systematics}.

\begin{table}[!ht]
\centering
\topcaption{%
    Summary of systematic uncertainty sources in the signal and background predictions.
    Electron- and muon-related uncertainties are listed together, with the values before and inside the parentheses referring to electrons and muons, respectively.
}
\label{tab:systematics}
\renewcommand{\arraystretch}{1.2}
\begin{tabular}{lcc}
    Source & Type & Uncertainty [\%] \\
    \hline
    \multicolumn{3}{c}{\textit{Signal prediction}\Large\strut} \\
    NNLO $K$-factor & Normalization & 4 \\
    Integrated luminosity & Normalization & 1.6 \\
    Pileup modeling & Shape & 4.6 \\
    \Pe (\PGm) trigger efficiency & Shape & 1 ($<$1) \\
    Prompt \Pe (\PGm) selection efficiency & Shape & 2--4 (1--3)  \\
    Nonprompt \Pe (\PGm) selection efficiency & Shape & 1--20 ($<$1) \\
    Tracking efficiency & Shape & 7.3 \\
    Jet energy scale \& resolution & Shape & 1--2 \\
    PFN score & Normalization & 10 \\[\cmsTabSkip]
    \multicolumn{3}{c}{\textit{Background prediction}} \\
    CR closure & Shape & 20--30 \\
    DY scale factor (OS \Cmm, low mass) & Shape & 20--50 \\
\end{tabular}
\end{table}

The $K$-factor applied to the normalization of HNL signal events has an uncertainty of 4\%, evaluated by performing variations of the factorization and renormalization scales and of the PDFs in the NNLO \PW boson production calculation.
Additionally, the \pt spectrum of the simulated \PW boson in SM \Wjets production at either NLO or LO accuracy was compared, and all differences were found to be covered by the $K$-factor uncertainty.

The integrated luminosity of the three data-taking years is measured with uncertainties between 1.2 and 2.5\%~\cite{CMS:LUM-17-003, CMS:LUM-17-004, CMS:LUM-18-002}, with a total uncertainty of 1.6\% for the combined data set.
The modeling uncertainty of pileup interactions is assessed by varying the total \pp inelastic cross section in simulation by $\pm$4.6\%.

The efficiencies of the trigger and prompt-lepton selections are measured in data and simulation with the ``tag-and-probe'' method applied to $\PZ\to\Pell^+\Pell^-$ events~\cite{CMS:EWK-10-002}.
Differences between efficiencies in data and simulation are corrected for with scale factors, and the uncertainties in these scale factors are considered for the HNL signal yield.
The uncertainties in the trigger scale factors are found to be about 1\% (less than 1\%) for the single-electron (single-muon) trigger, and in the selection scale factors to be 2--4 (1--3)\% for electrons (muons).

{\tolerance=800
The selection efficiency for nonprompt electrons is studied in asymmetric photon conversions events, following the procedure described in Ref.~\cite{CMS:EXO-20-009}.
Events with $\PZ\to\Pell^+\Pell^-\PGg\to\Pell^+\Pell^-\Pepm(\Pemp)$, where $(\Pemp)$ is a low-\pt electron that is not reconstructed in the detector, are selected.
Data-to-simulation differences are evaluated as a function of a displacement variable, with different results found for the different years of data taking, corresponding to differences in the material budget and its modeling in the detector simulation.
Systematic uncertainties in the range 1--20\% are found, largest for small displacements and depending on the data-taking year.
\par}

The nonprompt-muon selection efficiency is studied with a tag-and-probe method in $\PBpm\to\JPsi\PKpm\to\MM\PKpm$ events, where the muon pair from the \JPsi meson decay provides a good proxy to nonprompt muons in HNL decays~\cite{CMS:EXO-20-009}.
The differences between the efficiencies measured in data and simulated events are found to be small, with a systematic uncertainty of less than 1\%.

The track and SV reconstruction efficiency is studied using \PKzS meson decays to two charged particles, which provide an event signature with two displaced tracks originating from a common vertex~\cite{CMS:DP-2024-010}.
We find that reconstruction inefficiencies originate almost entirely from the track reconstruction inefficiency, while the SV reconstruction is almost 100\% efficient.
Scale factors are applied as a function of \pt and displacement to correct the simulation for differences with data, and we assign half the difference from unity as a systematic uncertainty.

Uncertainties in the momentum scale and resolution of prompt leptons~\cite{CMS:EGM-17-001, CMS:MUO-16-001} are found to be negligible.
Possible discrepancies between data and simulation in the momentum scale and resolution of nonprompt tracks are assessed by comparing the mean values and standard deviations of Gaussian fits to the reconstructed \PKzS candidate mass, in bins of SV displacement.
All discrepancies are found to be below 1\% and are neglected.

The systematic uncertainties associated with the jet energy scale and resolution~\cite{CMS:JME-13-004} are evaluated by independently scaling the energy of jets up and down by their uncertainty.
These variations result in a variation of the HNL signal yield by 1--2\%.

The PFN score validation in \PKzS meson decays, as described in Section~\ref{sec:PFN}, is used to evaluate a systematic uncertainty that covers differences between data and simulation in the PFN score for HNL events.
For high values of the PFN score, discrepancies between data and simulation of up to 10\% are observed, consistent for different PFN trainings and data-taking years.
Thus, we assign an overall 10\% uncertainty in the HNL yield.

The predicted background yields in the target region are subject to statistical uncertainties in the observed event yields in the sideband regions, resulting in uncertainties of up to 50\% in the target region yields.
Additional systematic uncertainties are assigned, as discussed in Section~\ref{sec:background} and described in the following.
From the agreement between predicted and observed yields in the CR, an uncertainty of 20--30\% in the background prediction is considered, separately for each channel.
For the additional scale factor applied to the background prediction in the OS \Cmm low-mass channel, the limited size of the DY MC sample used for the derivation of this scale factor results in an uncertainty of 20--50\%, depending on the SR bin.

\section{Results}
\label{sec:results}

\begin{figure}[!tp]
\centering
\includegraphics[width=0.95\textwidth]{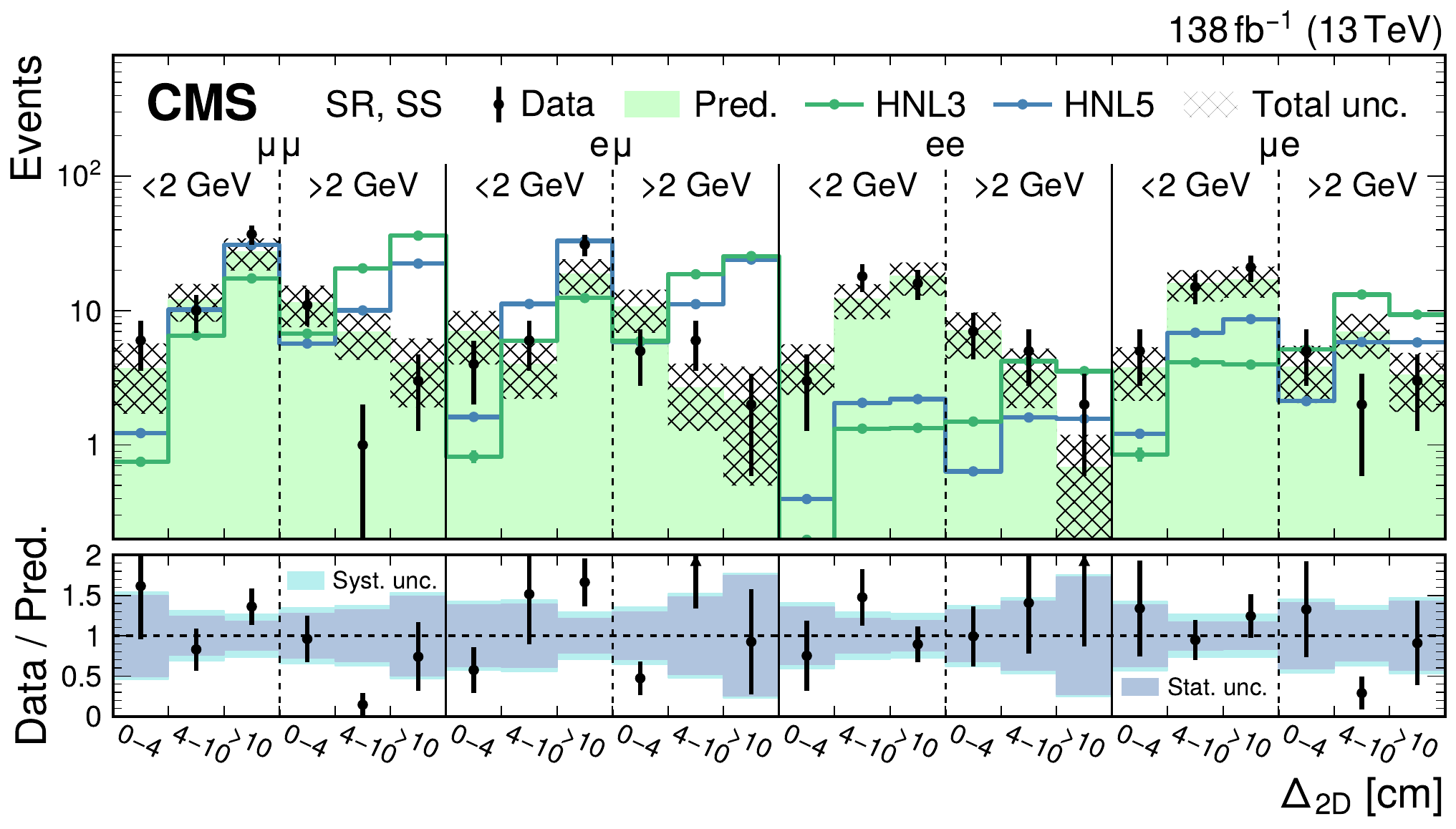}%
 \\[1ex]
 \includegraphics[width=0.95\textwidth]{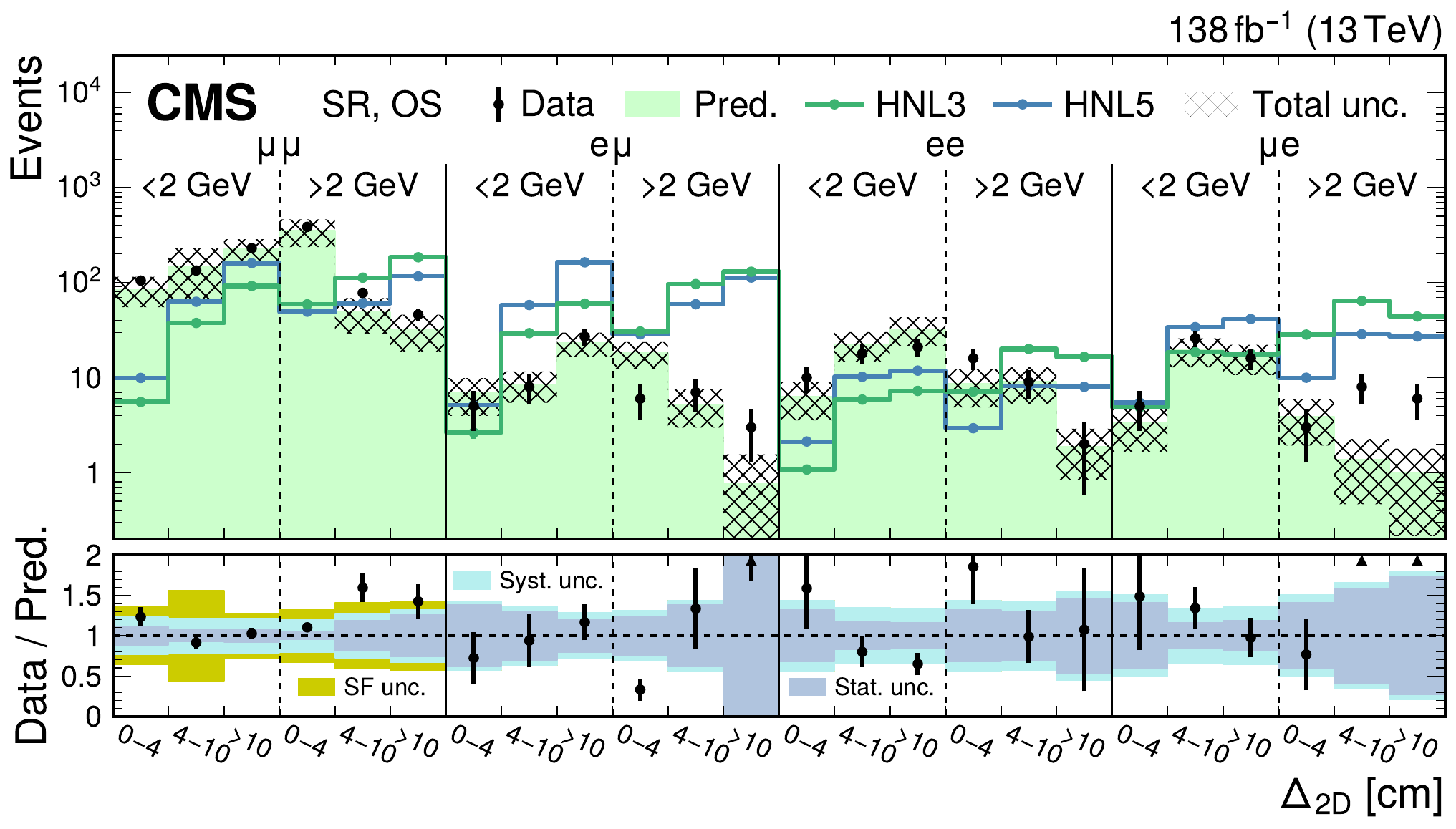}%
\caption{%
    Predicted and observed SR event yields for the SS (upper) and OS (lower) categories of the low-mass PFNs, binned by flavor channel, \mSV (as specified below the flavor channel), and \Deltwod.
    The hashed band represents the total systematic and statistical uncertainty in the background prediction.
    Signal predictions are shown for two HNL production hypotheses, with $\mhnl=3\GeV$ and $\mixparsq=9.9\times10^{-5}$ (HNL3), and $\mhnl=5\GeV$ and $\mixparsq=1.6\times10^{-6}$ (HNL5).
    The lower panels show the data-to-prediction ratio and the background prediction uncertainty is split into statistical (``stat.''), systematic (``syst.''), and DY scale factor (``SF'') contributions.
}
\label{fig:SRlow}
\end{figure}

\begin{figure}[!tp]
\centering
\includegraphics[width=0.95\textwidth]{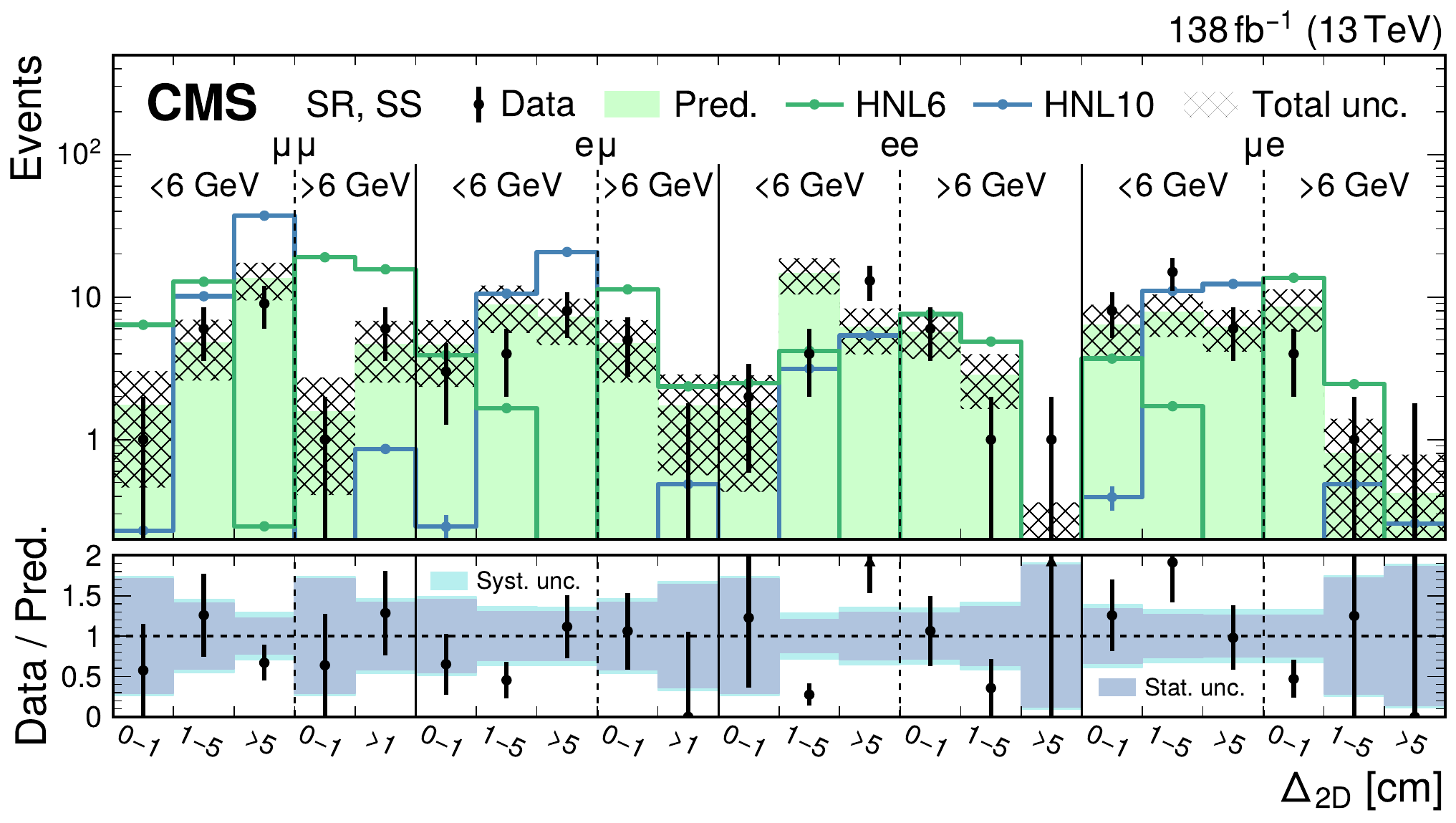}%
 \\[1ex]
\includegraphics[width=0.95\textwidth]{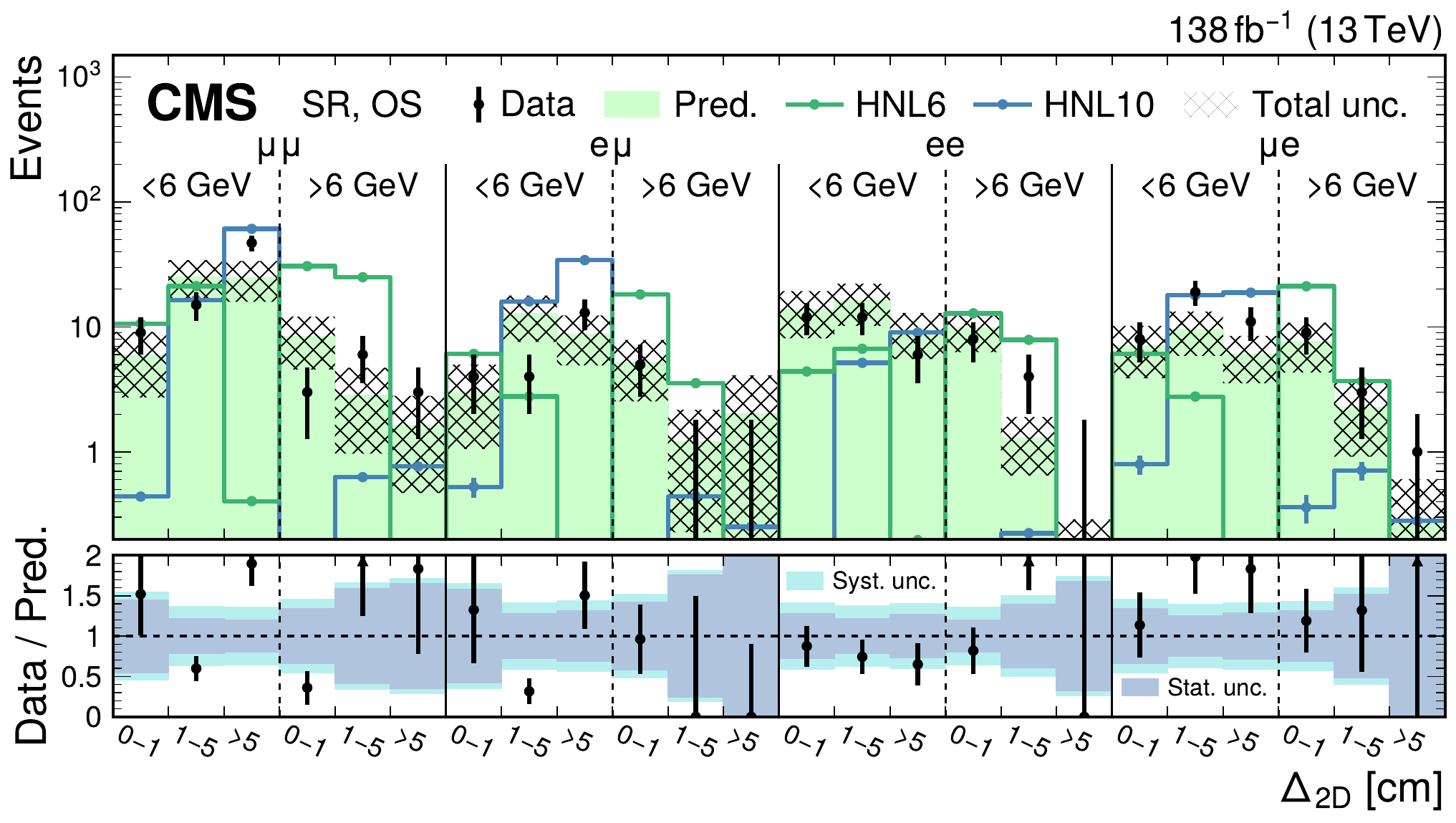}%
\caption{%
    Predicted and observed SR event yields for the SS (upper) and OS (lower) categories of the high-mass PFNs, binned by flavor channel, \mSV (as specified below the flavor channel), and \Deltwod.
    The hashed band represents the total systematic and statistical uncertainty in the background prediction.
    Signal predictions are shown for two HNL production hypotheses, with $\mhnl=6\GeV$ and $\mixparsq=2.0\times10^{-7}$ (HNL6), and $\mhnl=10\GeV$ and $\mixparsq=5.7\times10^{-7}$ (HNL10).
    The lower panels show the data-to-prediction ratio and the background prediction uncertainty is split into statistical (``stat.'') and systematic (``syst.'') contributions.
}
\label{fig:SRhigh}
\end{figure}

The yields in the four flavor channels are analyzed with both the low- and high-mass PFNs, and binned according to \mSV and \Deltwod, as defined in Section~\ref{sec:background}.
The observed data and the predicted background yields from the ABCD method in these bins are shown in Fig.~\ref{fig:SRlow} (\ref{fig:SRhigh}) for the low-mass (high-mass) PFNs.
The number of observed events in data is in agreement with the SM background expectations within the statistical and systematic uncertainties.
No significant data excess is found in any final state or SR bin.

For each $(\mhnl,\mixparsq)$ HNL signal scenario, we evaluate upper limits on the production cross section using the modified frequentist \CLs criterion~\cite{Junk:1999kv, Read:2002hq}.
A profile likelihood ratio is used as test statistic~\cite{CMS:NOTE-2011-005}, employing the asymptotic approximation~\cite{Cowan:2010js} in the limit setting procedure.
The binned likelihood is constructed from the observed data yields and the expected signal and background yields in the $(\mSV, \Deltwod)$ bins.
The expected signal yields are estimated from simulation.
The ABCD method for the estimation of the background yields discussed in Section~\ref{sec:background} is implemented in the likelihood similar to the method described in Ref.~\cite{CMS:EXO-21-013}: both the target and the three sideband regions of the SR are included, with possible signal contamination in the sideband regions explicitly considered, background templates constructed from the prefit difference between data and expected signal yields in the sideband regions, and nuisance parameters introduced to constrain the background yield in the target region to the ABCD estimate.
Further constrained nuisance parameters are introduced for the systematic uncertainties, which include effects on the shapes of the distributions.
The binned profile likelihood fits are performed with the CMS statistical analysis tool \textsc{combine}~\cite{CMS:CAT-23-001}, which is based on the \textsc{RooFit}~\cite{Verkerke:2003ir} and \textsc{RooStats}~\cite{Moneta:2010pm} frameworks.

\begin{figure}[!p]
\centering
\includegraphics[width=0.45\textwidth]{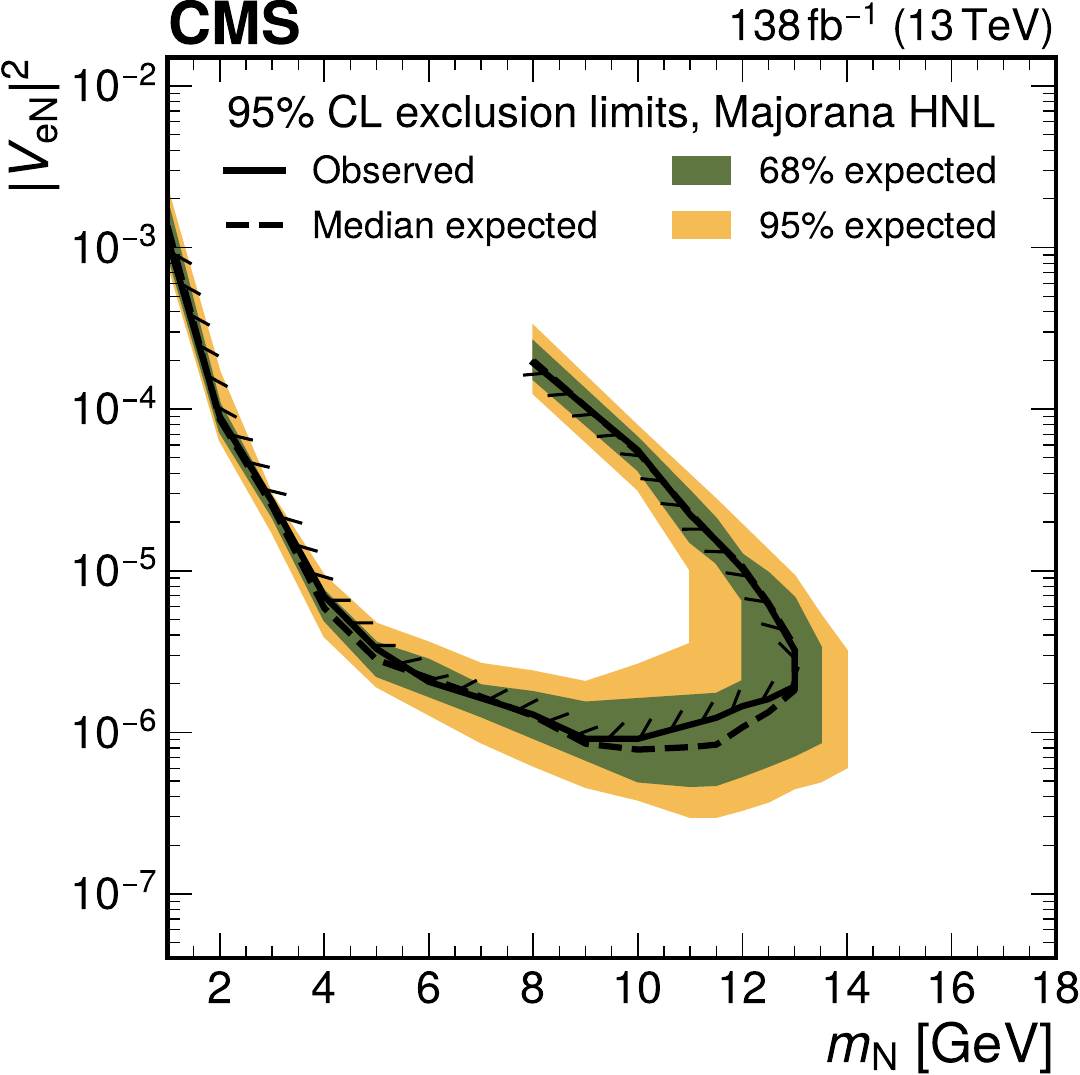}%
\hspace*{0.05\textwidth}%
\includegraphics[width=0.45\textwidth]{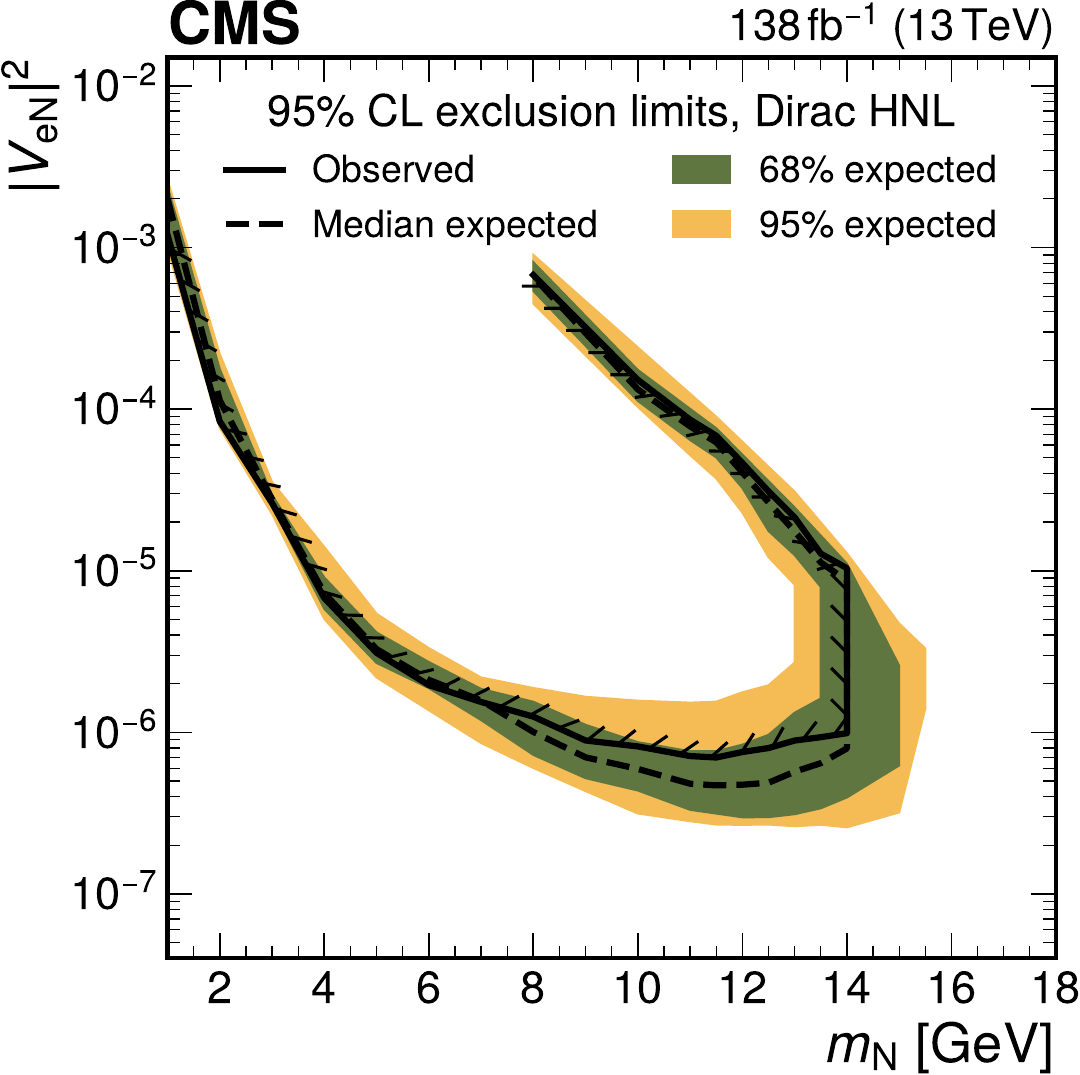}%
\\[1ex]
\includegraphics[width=0.45\textwidth]{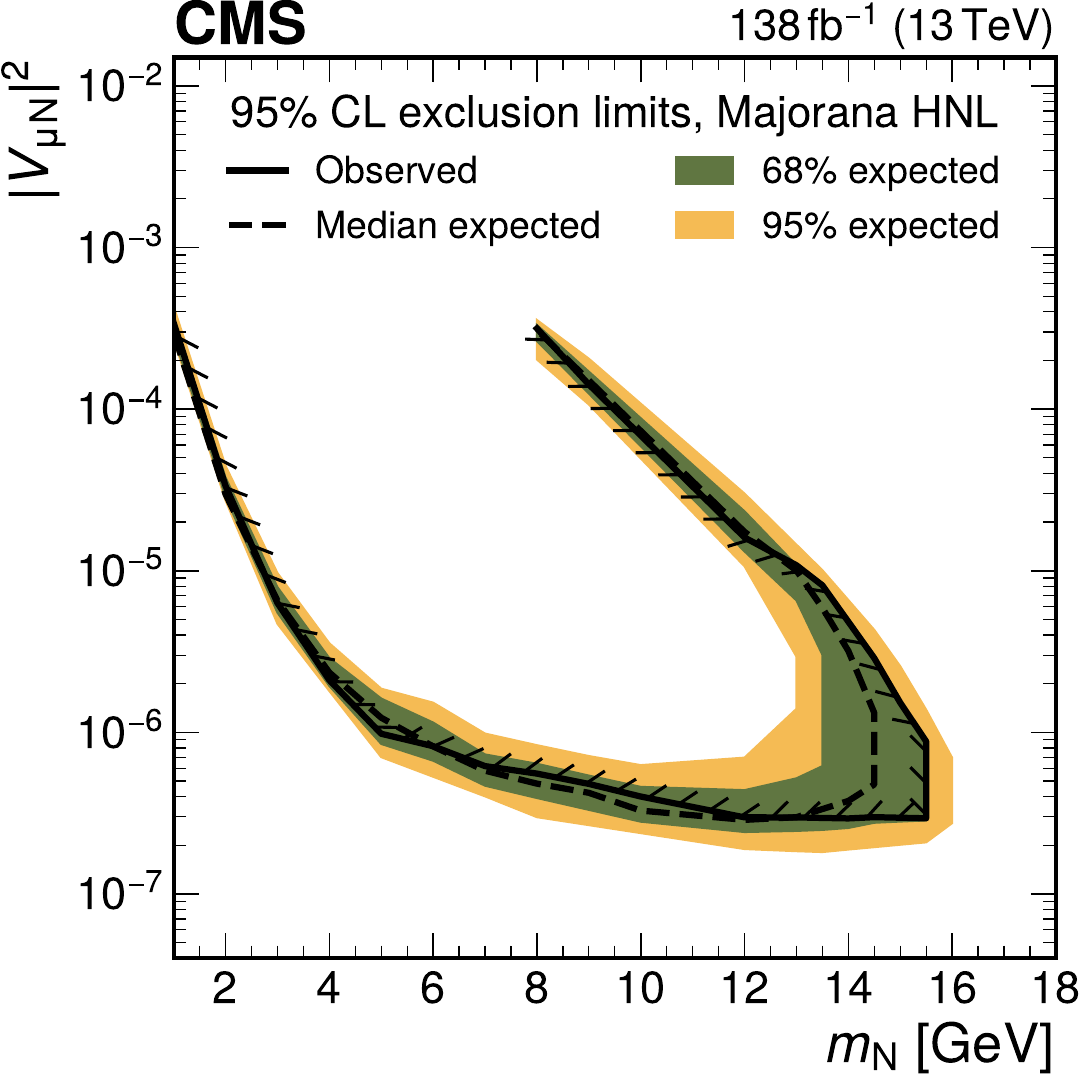}%
\hspace*{0.05\textwidth}%
\includegraphics[width=0.45\textwidth]{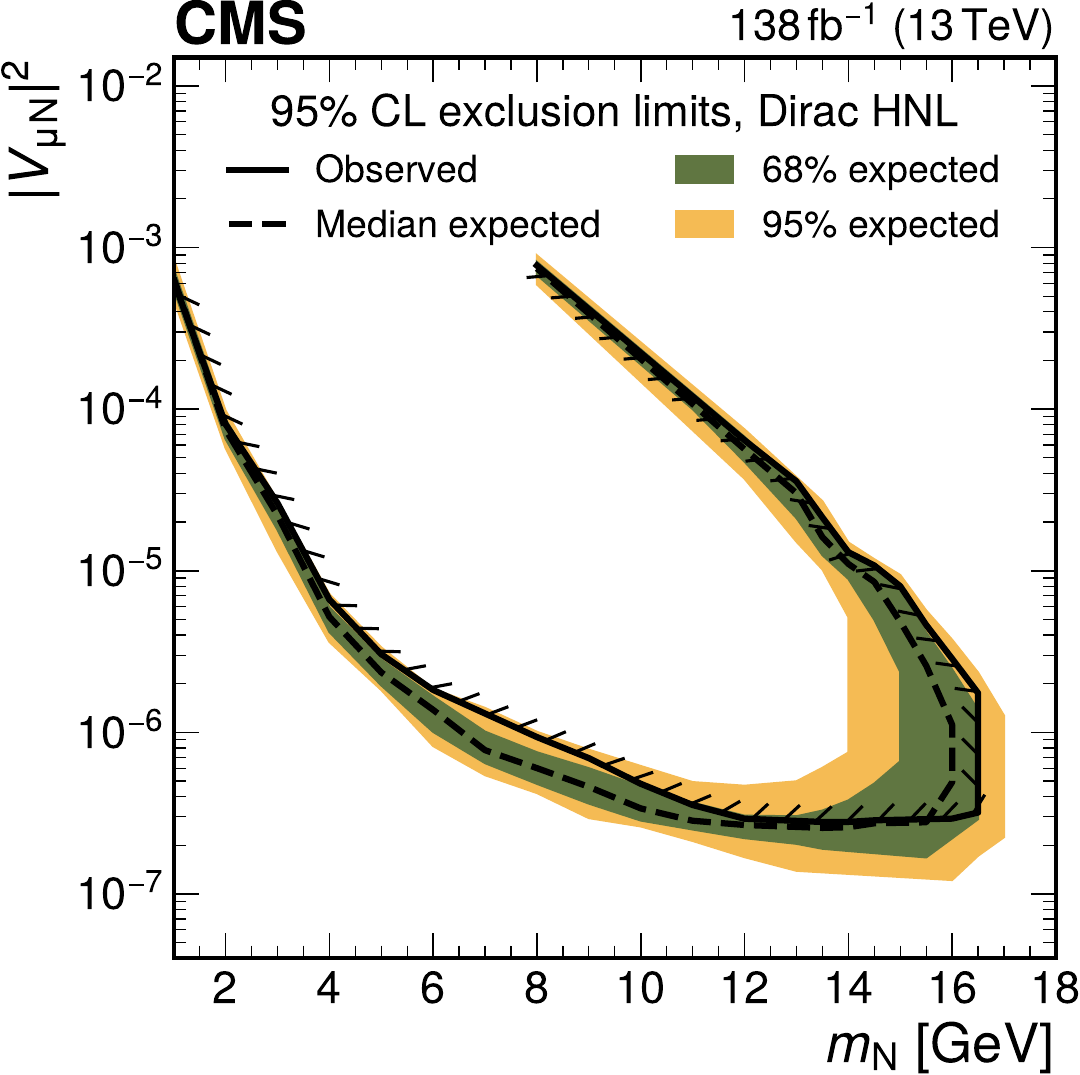}%
\\[1ex]
\includegraphics[width=0.45\textwidth]{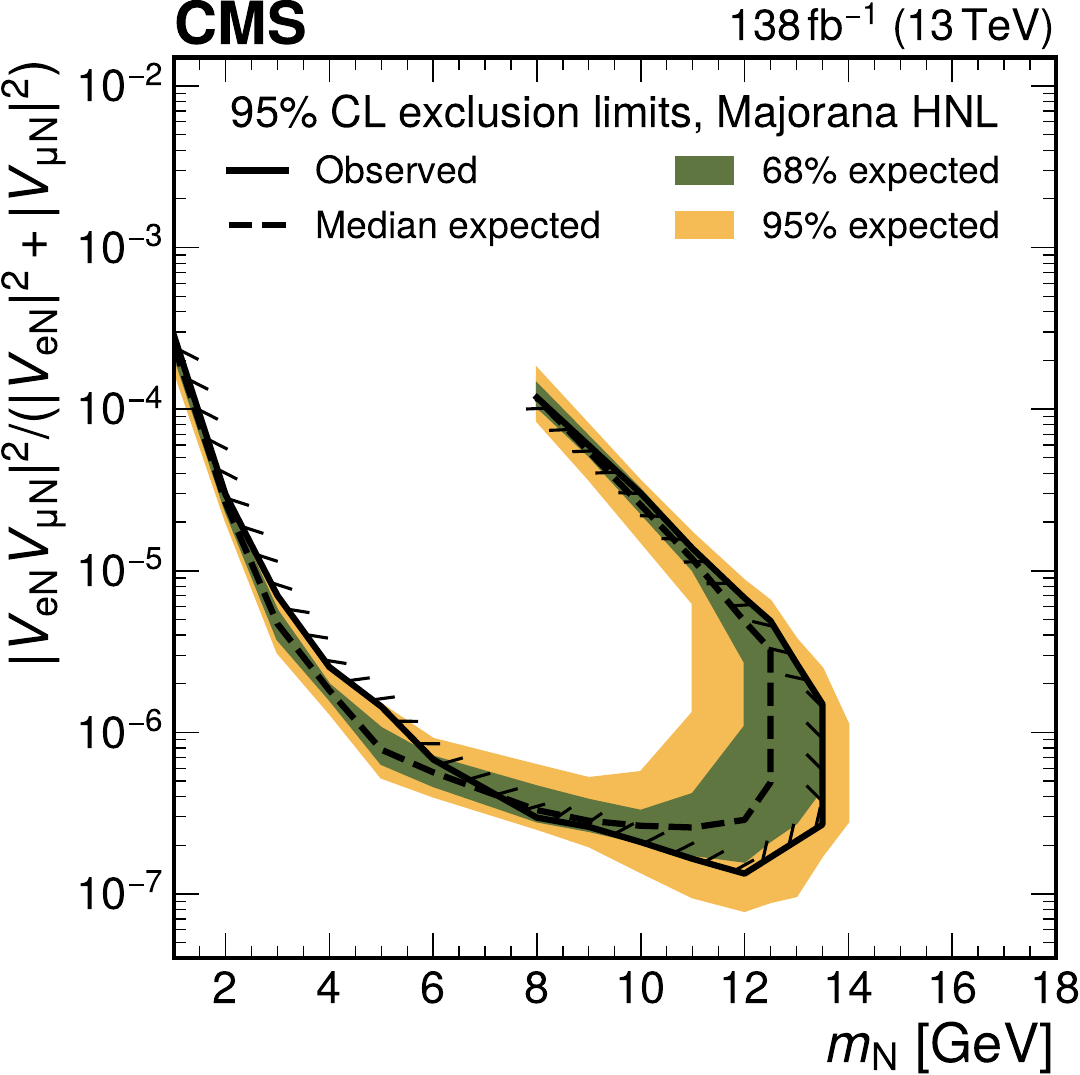}%
\hspace*{0.05\textwidth}%
\includegraphics[width=0.45\textwidth]{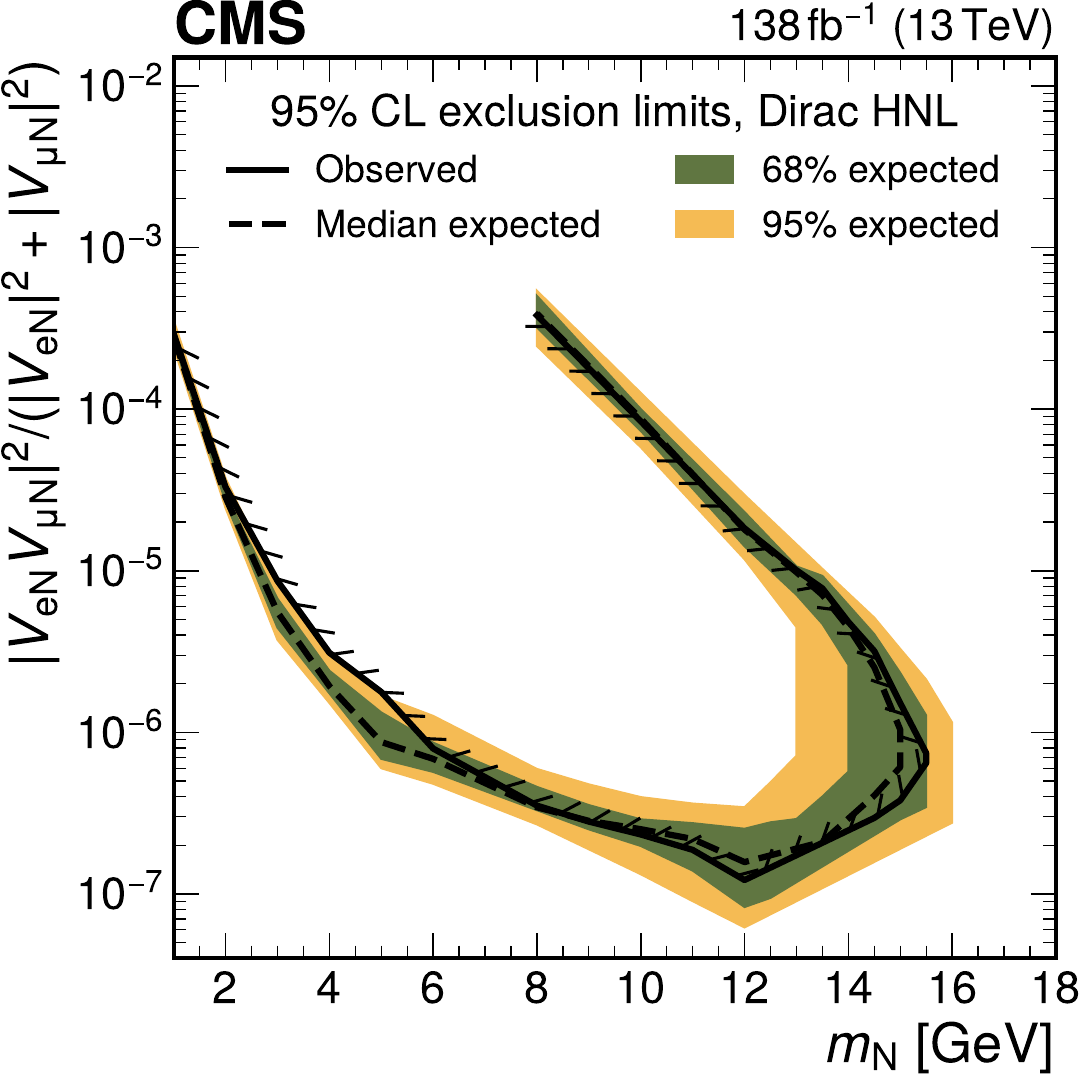}%
\caption{%
    Exclusion limits at 95\% \CL on \mixparesq (upper row), \mixparmsq (middle row), and \mixparemsq (lower row) as functions of \mhnl for a Majorana (left) and Dirac (right) HNL.
    The solid (dashed) black curve indicates the observed (expected) exclusion, where the parameter combinations inside the curve are excluded.
}
\label{fig:limits}
\end{figure}

We derive exclusion limits on \mixparesq, \mixparmsq, and \mixparemsq as functions of \mhnl, separately for the cases of Majorana and Dirac HNLs, using a grid of points in the $(\mhnl,\mixparsq)$ parameter space.
A signal scenario is excluded if the predicted cross section is incompatible with the observed data at 95\% confidence level (\CL).
The obtained limits are connected with straight lines between neighboring mass points.
We consider values of $1<\mhnl<20\GeV$ with a step size of 0.5 or 1\GeV.
For each mass point, per-event reweighting is employed to interpolate across the entire range of interest in \mixparsq, using all available MC samples, as described in Section~\ref{sec:samples}.
Generated samples with $\mixpare=\mixparm$ are used for deriving limits in the \Cem and \Cme channels, but since the HNL production cross section in these channels depends only on \mixparemsq~\cite{Atre:2009rg}, the results are applicable also to scenarios with different $\mixparesq:\mixparmsq$ ratios.
The obtained exclusion limits are presented in Fig.~\ref{fig:limits}.

The exclusion limits exhibit both an upper and a lower curve, corresponding to HNL scenarios with short and long lifetimes.
The upper curves are evaluated only for $\mhnl\geq8\GeV$; the prompt scenarios with lower \mhnl and high \mixparsq values are not well described by the available HNL signal samples and are already excluded by previous searches, \eg, by the results of the DELPHI Collaboration~\cite{DELPHI:1996qcc}.
The results of this study extend existing exclusion limits into the parameter space characterized by longer lifetimes and, correspondingly, smaller \mixparsq values.
In Table~\ref{tab:limit}, the observed limits are summarized for those mass points where the lowest values of \mixparsq are excluded per scenario.

\begin{table}[!ht]
\centering
\topcaption{%
    Comparison of lowest and highest \mixparsq values excluded at 95\% \CL for Majorana and Dirac HNLs with different coupling scenarios.
    For each scenario, the \mhnl value where the lowest \mixparsq value is excluded is shown.
}
\label{tab:limit}
\renewcommand{\arraystretch}{1.2}
\begin{tabular}{ccccc}
    \multirow{2}{*}{Type} & \multirow{2}{*}{Coupling} & \mhnl & \multicolumn{2}{c}{95\% \CL exclusion} \\[-2pt]
    & & [{\GeVns}] & Lowest value & Highest value \\
    \hline
    Majorana & \mixparesq & 10 & $9.1\times10^{-7}$ & $5.6\times10^{-5}$ \\
    Majorana & \mixparmsq & 14 & $2.9\times10^{-7}$ & $4.8\times10^{-6}$ \\
    Majorana & \mixparemsq & 12 & $1.3\times10^{-7}$ & $6.7\times10^{-6}$  \\
    Dirac & \mixparesq & 11.5 & $6.9\times10^{-7}$ & $6.8\times10^{-5}$ \\
    Dirac & \mixparmsq & 13.5 & $2.7\times10^{-7}$ & $2.1\times10^{-5}$ \\
    Dirac & \mixparemsq & 12 & $1.2\times10^{-7}$ & $1.8\times10^{-5}$ \\
\end{tabular}
\end{table}

Other experiments and searches have established more stringent limits for some of the mass range probed in this analysis.
The CHARM experiment~\cite{CHARM:1985nku} has excluded electron (muon) couplings for $1.0<\mhnl<2.1$ (1.9)\GeV, with \mixparsq values as low as $10^{-7}$.
The previous CMS search for HNLs decaying in the muon system~\cite{CMS:EXO-22-017} found more stringent limits in the mass range 1--2.8\GeV, with electron (muon) couplings excluded with \mixparsq values as low as 0.9~(0.5)${}\times10^{-6}$.
The displaced trilepton search~\cite{CMS:EXO-20-009} and the displaced-jet-tagger analysis in the same final state~\cite{CMS:EXO-21-013} probe similar \mhnl and \mixparsq regions as this work.
In the mass range 2--10\GeV, the two searches exhibit sensitivities that are comparable to our results.
For masses of 10--17\GeV and \mixparsq values below $10^{-5}$, the requirement of an SV in this analysis results in the removal of a significant amount of background from the SRs, resulting in more stringent limits when compared to Ref.~\cite{CMS:EXO-21-013}.
For masses above 10\GeV and higher coupling values (\ie, shorter lifetime), however, the displaced-jet-tagger approach shows a better sensitivity.
Compared to the recent prompt trilepton search~\cite{CMS:EXO-22-011}, the limits on \mixparesq and \mixparmsq presented here are more stringent in the probed mass range by more than one order of magnitude.

\section{Summary}

A search for long-lived heavy neutral leptons (HNLs) has been presented using proton-proton collision events with one prompt lepton and a system of a nonprompt lepton and a jet associated with a secondary vertex.
The data set corresponds to 138\fbinv and was collected by the CMS experiment at the LHC in 2016--2018.
Secondary vertices are reconstructed and matched with nonprompt leptons to find candidate vertices for the HNL decay, and a dedicated machine-learning method is utilized to distinguish between secondary vertices associated with HNL decays and those from background sources.
No excess of events above the standard model background prediction obtained from control samples in data is found.
Exclusion limits at 95\% confidence level are evaluated for different HNL coupling scenarios as functions of the HNL mass and the mixing parameter with standard model neutrinos.
The obtained exclusion limits cover HNL masses from 1 to 16.5\GeV and squared mixing parameters as low as $2\times10^{-7}$, depending on the scenario.
These results exceed previous experimental constraints derived in the single-lepton decay channel in the mass range 11--16.5\GeV.
For some of the considered coupling scenarios and mass ranges, the presented limits are the strongest to date.

\begin{acknowledgments}
    We congratulate our colleagues in the CERN accelerator departments for the excellent performance of the LHC and thank the technical and administrative staffs at CERN and at other CMS institutes for their contributions to the success of the CMS effort. In addition, we gratefully acknowledge the computing centers and personnel of the Worldwide LHC Computing Grid and other centers for delivering so effectively the computing infrastructure essential to our analyses. Finally, we acknowledge the enduring support for the construction and operation of the LHC, the CMS detector, and the supporting computing infrastructure provided by the following funding agencies: SC (Armenia), BMBWF and FWF (Austria); FNRS and FWO (Belgium); CNPq, CAPES, FAPERJ, FAPERGS, and FAPESP (Brazil); MES and BNSF (Bulgaria); CERN; CAS, MoST, and NSFC (China); MINCIENCIAS (Colombia); MSES and CSF (Croatia); RIF (Cyprus); SENESCYT (Ecuador); ERC PRG, RVTT3 and MoER TK202 (Estonia); Academy of Finland, MEC, and HIP (Finland); CEA and CNRS/IN2P3 (France); SRNSF (Georgia); BMBF, DFG, and HGF (Germany); GSRI (Greece); NKFIH (Hungary); DAE and DST (India); IPM (Iran); SFI (Ireland); INFN (Italy); MSIP and NRF (Republic of Korea); MES (Latvia); LMTLT (Lithuania); MOE and UM (Malaysia); BUAP, CINVESTAV, CONACYT, LNS, SEP, and UASLP-FAI (Mexico); MOS (Montenegro); MBIE (New Zealand); PAEC (Pakistan); MES and NSC (Poland); FCT (Portugal); MESTD (Serbia); MCIN/AEI and PCTI (Spain); MOSTR (Sri Lanka); Swiss Funding Agencies (Switzerland); MST (Taipei); MHESI and NSTDA (Thailand); TUBITAK and TENMAK (Turkey); NASU (Ukraine); STFC (United Kingdom); DOE and NSF (USA).

    \hyphenation{Rachada-pisek} Individuals have received support from the Marie-Curie program and the European Research Council and Horizon 2020 Grant, contract Nos.\ 675440, 724704, 752730, 758316, 765710, 824093, 101115353, 101002207, and COST Action CA16108 (European Union); the Leventis Foundation; the Alfred P.\ Sloan Foundation; the Alexander von Humboldt Foundation; the Science Committee, project no. 22rl-037 (Armenia); the Belgian Federal Science Policy Office; the Fonds pour la Formation \`a la Recherche dans l'Industrie et dans l'Agriculture (FRIA-Belgium); the F.R.S.-FNRS and FWO (Belgium) under the ``Excellence of Science -- EOS" -- be.h project n.\ 30820817; the Beijing Municipal Science \& Technology Commission, No. Z191100007219010 and Fundamental Research Funds for the Central Universities (China); the Ministry of Education, Youth and Sports (MEYS) of the Czech Republic; the Shota Rustaveli National Science Foundation, grant FR-22-985 (Georgia); the Deutsche Forschungsgemeinschaft (DFG), among others, under Germany's Excellence Strategy -- EXC 2121 ``Quantum Universe" -- 390833306, and under project number 400140256 - GRK2497; the Hellenic Foundation for Research and Innovation (HFRI), Project Number 2288 (Greece); the Hungarian Academy of Sciences, the New National Excellence Program - \'UNKP, the NKFIH research grants K 131991, K 133046, K 138136, K 143460, K 143477, K 146913, K 146914, K 147048, 2020-2.2.1-ED-2021-00181, and TKP2021-NKTA-64 (Hungary); the Council of Science and Industrial Research, India; ICSC -- National Research Center for High Performance Computing, Big Data and Quantum Computing and FAIR -- Future Artificial Intelligence Research, funded by the NextGenerationEU program (Italy); the Latvian Council of Science; the Ministry of Education and Science, project no. 2022/WK/14, and the National Science Center, contracts Opus 2021/41/B/ST2/01369 and 2021/43/B/ST2/01552 (Poland); the Funda\c{c}\~ao para a Ci\^encia e a Tecnologia, grant CEECIND/01334/2018 (Portugal); the National Priorities Research Program by Qatar National Research Fund; MCIN/AEI/10.13039/501100011033, ERDF ``a way of making Europe", and the Programa Estatal de Fomento de la Investigaci{\'o}n Cient{\'i}fica y T{\'e}cnica de Excelencia Mar\'{\i}a de Maeztu, grant MDM-2017-0765 and Programa Severo Ochoa del Principado de Asturias (Spain); the Chulalongkorn Academic into Its 2nd Century Project Advancement Project, and the National Science, Research and Innovation Fund via the Program Management Unit for Human Resources \& Institutional Development, Research and Innovation, grant B39G670016 (Thailand); the Kavli Foundation; the Nvidia Corporation; the SuperMicro Corporation; the Welch Foundation, contract C-1845; and the Weston Havens Foundation (USA).
\end{acknowledgments}

\bibliography{auto_generated}
\cleardoublepage \appendix\section{The CMS Collaboration \label{app:collab}}\begin{sloppypar}\hyphenpenalty=5000\widowpenalty=500\clubpenalty=5000
\cmsinstitute{Yerevan Physics Institute, Yerevan, Armenia}
{\tolerance=6000
A.~Hayrapetyan, A.~Tumasyan\cmsAuthorMark{1}\cmsorcid{0009-0000-0684-6742}
\par}
\cmsinstitute{Institut f\"{u}r Hochenergiephysik, Vienna, Austria}
{\tolerance=6000
W.~Adam\cmsorcid{0000-0001-9099-4341}, J.W.~Andrejkovic, T.~Bergauer\cmsorcid{0000-0002-5786-0293}, S.~Chatterjee\cmsorcid{0000-0003-2660-0349}, K.~Damanakis\cmsorcid{0000-0001-5389-2872}, M.~Dragicevic\cmsorcid{0000-0003-1967-6783}, P.S.~Hussain\cmsorcid{0000-0002-4825-5278}, M.~Jeitler\cmsAuthorMark{2}\cmsorcid{0000-0002-5141-9560}, N.~Krammer\cmsorcid{0000-0002-0548-0985}, A.~Li\cmsorcid{0000-0002-4547-116X}, D.~Liko\cmsorcid{0000-0002-3380-473X}, I.~Mikulec\cmsorcid{0000-0003-0385-2746}, J.~Schieck\cmsAuthorMark{2}\cmsorcid{0000-0002-1058-8093}, R.~Sch\"{o}fbeck\cmsorcid{0000-0002-2332-8784}, D.~Schwarz\cmsorcid{0000-0002-3821-7331}, M.~Sonawane\cmsorcid{0000-0003-0510-7010}, S.~Templ\cmsorcid{0000-0003-3137-5692}, W.~Waltenberger\cmsorcid{0000-0002-6215-7228}, C.-E.~Wulz\cmsAuthorMark{2}\cmsorcid{0000-0001-9226-5812}
% ...
\par}
\vskip\cmsinstskip
\dag:~Deceased\\
$^{1}$Also at Yerevan State University, Yerevan, Armenia\\
$^{2}$Also at TU Wien, Vienna, Austria\\
% ...
\end{sloppypar}
%%% END EDITABLE REGION %%%
% skeleton_end
\end{document}